\documentclass[%
reprint, 
superscriptaddress,
 amsmath,amssymb,
 aps,
]{revtex4-2}
\usepackage{array}[=2016-10-06]
\usepackage{tabularx}
\usepackage[T1]{fontenc}
\usepackage{multirow} 
\usepackage{graphicx}
\usepackage{dcolumn}
\usepackage{bm}
\usepackage{siunitx}
\usepackage[mathlines]{lineno}
\usepackage[dvipsnames]{xcolor}
\usepackage{nicematrix}
\usepackage{ulem} 
\usepackage{layouts}
\usepackage{xr-hyper}
\usepackage{hyperref} 
\hypersetup{
    colorlinks,
    linkcolor={blue!50!black},
    citecolor={blue!50!black},
    urlcolor={blue!80!black}
} 
\let\micro\relax

\newcommand{\nocontentsline}[3]{}
\newcommand\stoptoc{%
   \let\origcontentsline\addcontentsline
   \let\addcontentsline\nocontentsline
}
\newcommand\resumetoc{%
   \let\addcontentsline\origcontentsline
}

\usepackage{subfiles} 

\begin{document}
\stoptoc

\preprint{APS/123-QED}

\title{Coherence of a hole spin flopping-mode qubit in a circuit quantum electrodynamics environment}



\author{Léo Noirot}
\affiliation{Univ. Grenoble Alpes, CEA, Grenoble INP, IRIG-Pheliqs, Grenoble, France.}
\author{Cécile X. Yu}
\affiliation{Univ. Grenoble Alpes, CEA, Grenoble INP, IRIG-Pheliqs, Grenoble, France.}
\affiliation{QuTech and Kavli Institude of Nanoscience, Delft University of Technology, P.O. Box 2046, Delft, 2600 GA, Delft, The Netherlands.}
\author{José C. Abadillo-Uriel}
\affiliation{Univ. Grenoble Alpes, CEA, IRIG-MEM-L\_Sim, Grenoble, France.}
\affiliation{Instituto de Ciencia de Materiales de Madrid, Consejo Superior de Investigaciones Cientificas, Madrid 28049, Spain.}
\author{Étienne Dumur}
\affiliation{Univ. Grenoble Alpes, CEA, Grenoble INP, IRIG-Pheliqs, Grenoble, France.}
\author{Heimanu Niebojewski}
\affiliation{Univ. Grenoble Alpes, CEA, LETI, Minatec Campus, Grenoble, France.}
\author{Benoit Bertrand}
\affiliation{Univ. Grenoble Alpes, CEA, LETI, Minatec Campus, Grenoble, France.}
\author{Romain Maurand}
\affiliation{Univ. Grenoble Alpes, CEA, Grenoble INP, IRIG-Pheliqs, Grenoble, France.}
\email{romain.maurand@cea.fr}
\author{Simon Zihlmann}
\affiliation{Univ. Grenoble Alpes, CEA, Grenoble INP, IRIG-Pheliqs, Grenoble, France.}

\date{\today}

\begin{abstract}
The entanglement of microwave photons and spin qubits in silicon represents an important step forward for quantum information processing utilizing semiconductor quantum dots. Such hybrid spin circuit quantum electrodynamics experiments have been achieved by delocalizing spins in a double quantum dot with spin-orbit interactions, producing a flopping-mode spin qubit with a substantial electric dipole moment. Unfortunately, demonstrations of these qubits have not shown the coherence properties necessary to be used as practical single qubits. Here, we present a flopping-mode hole spin qubit in a silicon nanowire coupled to a high-impedance niobium nitride microwave readout resonator. We report Rabi frequencies exceeding \SI{100}{MHz} with coherence times in the microsecond range, resulting in a single gate quality factor of 380. This establishes the speed and reliability of flopping-mode spin qubits. Moreover, using the large frequency tunability of the qubit, we find that radiative decay is the main relaxation channel in our experiment and argue that photon shot noise is the major source of dephasing. These results suggest that optimized microwave engineering can unlock the potential of flopping-mode spin qubits in hybrid circuit quantum electrodynamics architectures, offering a scalable and robust platform for fast and coherent spin qubits with strong coupling to microwave photons.
\end{abstract}

\maketitle


\section*{Introduction}
\begin{figure*}[htbp]
    \includegraphics[width=1\linewidth]{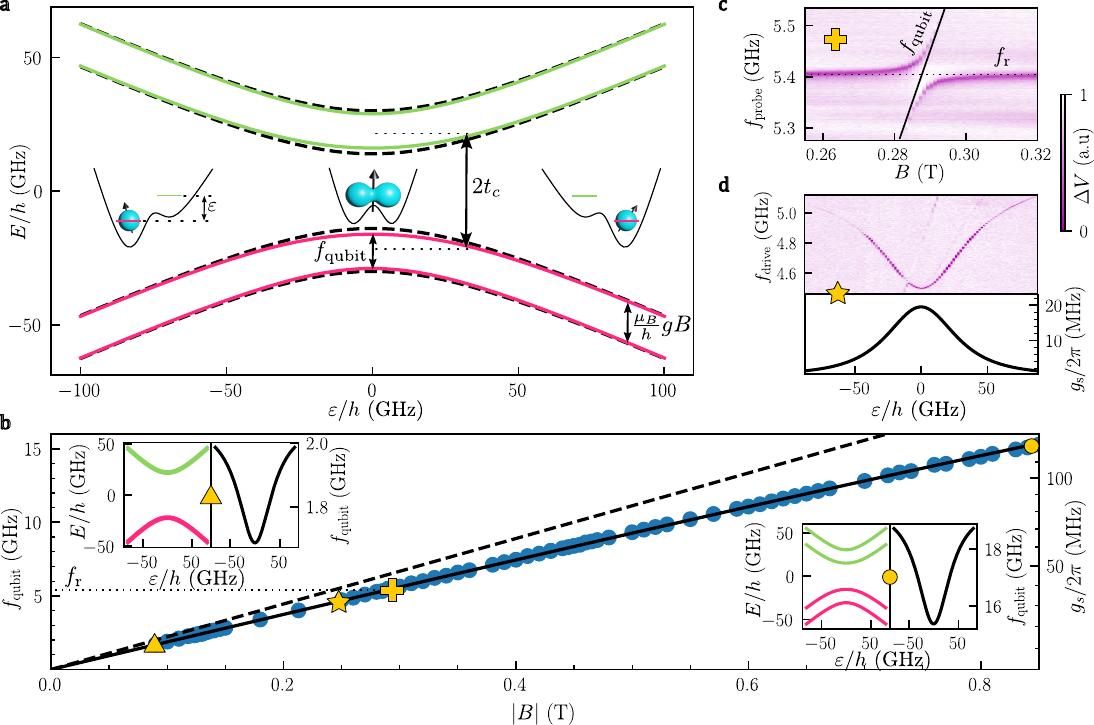}

   \caption{\textbf{Flopping-mode spin qubit:} (\textbf{a}) Energy level diagram of a spin in a DQD with strong SOI. The orbital molecular states: bonding (pink) and anti-bonding (green), are separated by the tunneling energy $2t_c$ and are spin-split under the action of the external magnetic field $B$. In the absence of SOI, the spin-splitting is given by the Zeeman energy (dashed-lines), whereas in the presence of SOI, the spin-charge mixing reduces the splitting around $\varepsilon=0$ (solid lines). The flopping-mode spin qubit is then defined between the spin-split levels of the orbital bonding state. The inset schematically illustrates the delocalization of the wave function as the QDs' orbitals are brought in resonance. 
   (\textbf{b}) Measurement by two-tone spectroscopy of $f_\mathrm{qubit}$ (blue dots) as a function of $B$ at $\varepsilon=0$. The solid line is a fit to the FM spin qubit Hamiltonian introduced in the supplementary section~\ref{sec:FM model}. The dashed line corresponds to the expected Zeeman energy in absence of SOI.
   Left (resp. right) inset shows the energy level diagram and FM spin transition energy at $B=\SI{0.09}{T}$ (resp. $B=\SI{0.85}{T}$). The secondary y-axis of the main plot shows the magnetic field dependence of $g_s$ as expected from the model ($g_s \propto B$). The yellow symbols pinpoint the magnetic field and $f_\mathrm{qubit}$ corresponding to (\textbf{c}), (\textbf{d}) and to the insets of (\textbf{b}).  (\textbf{c}) Transmitted amplitude as a function of $B$ and of the probe frequency $f_\mathrm{probe}$ at $\varepsilon=0$. The avoided crossing reveals the vacuum Rabi splitting between the readout-resonator and the FM spin qubit allowing the extraction of $g_s$ at resonance ($f_\mathrm{qubit} = f_r$). (\textbf{d}) Two-tone spectroscopy of the FM spin qubit transition exhibiting a first-order detuning sweet-spot. The lower panel shows the corresponding evolution of $g_s$ in $\varepsilon$ illustrating the quenching of the electric-dipole as the spin gets localized in one quantum dot ($|\varepsilon|> 2 t_\mathrm{c}$).}
    \label{fig1}
\end{figure*}
Spin qubits in semiconductor quantum dots form a diverse and versatile family, distinguished by the number of spins and quantum dot sites used to encode quantum information \cite{BurkardReview}. The simplest realization, the single-spin qubit or Loss-DiVincenzo qubit, harnesses the spin degree of freedom of an electron or a hole confined in a single quantum dot \cite{Loss}. Two-electron singlet-triplet qubits encode quantum information in the spin state of a coupled pair, based on exchange interaction \cite{2005_Petta}. This concept can be further extended to three coupled spins that form an exchange-only qubit \cite{2017_Malinowski}. With pioneering works on GaAs and the blooming of silicon and germanium quantum dots, the spin qubit family has achieved key milestones for universal quantum logic~\cite{BurkardReview, 2022_Stano} and is now aiming for scaling \cite{2017_Vandersypen}. However, inherently short-range spin-spin coupling based on exchange interaction, together with increasingly dense wiring demands \cite{2023_Borsoi}, has constrained spin qubit arrays to relatively small sizes~\cite{ 2022_Philips, 2022_Takeda, 2024_Wang}. 

The key challenge now is the realization of long-range entanglement essential for scalable architectures \cite{2012_Fowler}. In this context, coupling spin qubits to microwave photons in superconducting resonators provides a promising route for long-distance spin-spin interaction, leveraging the well-developed toolbox of cQED. However, the weak magnetic dipole moment of a spin limits its direct interaction with microwave photons. Instead, strong spin-photon coupling typically relies on spin-charge hybridization, which grants an electric dipole to spin transitions. Spin qubits encoded in multiple spins and over multiple sites naturally come with an electric dipole originating from the exchange interaction and the spatially extended wave function. This has been harnessed to strongly couple resonant exchange \cite{2018_Landig} or singlet-triplet qubits \cite{2022_Boettcher, 2024_Ungerer} to microwave photons. In order to reduce the complexity with multiparticle encodings \cite{BurkardReview}, a single spin qubit with a large electric dipole is sought. As a single spin does not benefit from exchange interaction, spin-orbit interaction (SOI) instead may be leveraged for spin-charge hybridization. Therefore, de-localizing a single spin over two quantum dot sites in the presence of SOI, will create a so-called flopping-mode (FM) spin qubit~\cite{Childress2004, Cottet2010,Hu2012}. The demonstrations of strong spin-photon coupling for electrons~\cite{2018_Samkharadze, 2018_Mi} and holes~\cite{Yu} in silicon have relied on FM spin qubits, illustrating their large electric dipole and potential for long-range entanglement.

Although proof-of-principles of photon-mediated spin-spin interactions~\cite{2019_Borjans, 2022_Harvey-Collard} and iSWAP oscillations~\cite{2024_Dijkema} have been reported, their fidelities remain limited by the short coherence times observed for FM spin qubits. Decoherence mechanisms commonly affecting spin qubits, such as electrical noise \cite{Benito2019, Croot, 2024_Dijkema} and poor charge coherence properties \cite{Benito2019}, are frequently cited as potential contributors to these limitations. However, direct experimental evidence, pinpointing the exact coherence limits, remain elusive and it is not clear whether a FM spin qubit can exhibit high single qubit performances.

In this work, we conduct a comprehensive study of a hole spin FM qubit embedded in a cQED environment, a fundamental setting for future long-range entanglement experiments with spin qubits. Contrary to previous expectations \cite{2018_Samkharadze, 2018_Mi, Croot, Yu, 2024_Dijkema}, we reveal that decoherence in our device is dominated by light-matter interaction in the form of photon emission for relaxation and photon-shot noise for dephasing, rather than by mechanisms commonly limiting spin qubits. By mitigating these effects, we demonstrate echo dephasing times up to $\SI{5}{\micro s}$ and Rabi frequencies as high as $\SI{130}{MHz}$, potentially allowing for single-qubit fidelities of $\SI{99.9}{\%}$. With strong spin-photon coupling and promising single-qubit properties demonstrated here, hole spin FM qubits emerge as an interesting qubit candidate for quantum information processing and simulation based on 2D cQED integration~\cite{2017_Nigg}.
\section*{A single spin in a double quantum dot}
In this study, we delocalize a hole in a double quantum dot (DQD) formed in a natural silicon nanowire \cite{Yu}. Similarly to the hydrogen molecule, the hybridization of the QDs' orbitals forms a bonding and an antibonding state. Under static magnetic field, the spin degeneracy of these molecular orbitals is lifted, thus creating a four-level spin-orbit system, whose energy dispersion is depicted in Fig.~\ref{fig1}~(\textbf{a}) as a function of the energy detuning $\varepsilon$ between the two QDs. The two lowest states encode a spin qubit at frequency $f_\mathrm{qubit}$ associated with an orbital wave function spread over the entire DQD. In the presence of SOI, the orbital and spin degrees of freedom are mixed, granting a large electric dipole moment to the FM spin qubit \cite{BurkardReview}.

Here we connect a high-impedance NbN resonator of resonance frequency $f_\mathrm{r}$ \cite{2021_Yu} to one gate defining the DQD enabling strong spin-photon coupling, as previously demonstrated in \cite{Yu}. In the dispersive regime, when the frequency mismatch between the resonator and the qubit is larger than their coupling, the spin-photon interaction leads to a shift of the resonator frequency depending on the qubit state. This allows to determine the spin state by probing the resonator response in a microwave measurement without requiering a local charge sensor~\cite{2021_Blais}. 
Fig.~\ref{fig1}~(\textbf{b}) presents the measured magnetic field dependence of the qubit frequency when the hole wave function is completely delocalized between the two QD sites ($\varepsilon=0$). With the nanowire geometry leading to a particularly strong SOI \cite{2018_Kloeffel}, and a large tunnel coupling ($2t_\mathrm{c}=\SI{44}{GHz}$), Fig.~\ref{fig1}(\textbf{b}) demonstrates that $f_\mathrm{qubit}$ is tunable over more than an order of magnitude by the external magnetic field, while being readable through its dispersive interaction with the microwave resonator.

Unlike a pure spin transition, the FM spin transition energy differs from the expected Zeeman energy $g\mu_B B$ due to spin-charge hybridization, which renormalizes the g-factor $g$. The degree of spin-charge hybridization can be modeled by fitting the experimental data of Fig.~\ref{fig1}~(\textbf{b}) to a model that considers tunneling in the presence of SOI, giving insights into the key parameters of the FM qubit (see supplementary section \ref{sec:FM model}). From the model we obtain the spin-photon coupling  $g_{\rm{s}}$ for each qubit frequency as indicated in Fig.~\ref{fig1}~(\textbf{b}) and which is confirmed by the vacuum Rabi splitting shown in Fig.~\ref{fig1}~(\textbf{c}) when the FM qubit is resonant with the resonator ($f_\mathrm{qubit} = f_\mathrm{r}$). As a result of the g-factor renormalization, the qubit frequency exhibits a minimum at zero detuning resulting in a first-order sweet spot with respect to charge noise (detuning noise) as illustrated in the insets in Fig.~\ref{fig1}~(\textbf{b}) and confirmed by two-tone spectroscopy in Fig.~\ref{fig1}~(\textbf{d}).
Moving away from the sweet-spot, the hole becomes increasingly localized in a single quantum dot when increasing $|\varepsilon|$, leading to a rapid quench of the electric dipole and consequently to a vanishing spin-photon coupling, see Fig.~\ref{fig1}~(\textbf{d}) bottom panel. To summarize, the FM spin qubit exhibits by construction a reciprocal sweetness \cite{2023_Michal, 2024_Bassi} consisting in a first-order sweet-spot with respect to $\varepsilon$ noise associated with a maximal electric dipole \cite{Benito2019,Croot}. This property should lead to high-quality single qubit performances but has never been demonstrated so far.

\begin{figure}[htbp]
    \includegraphics[width=\columnwidth]{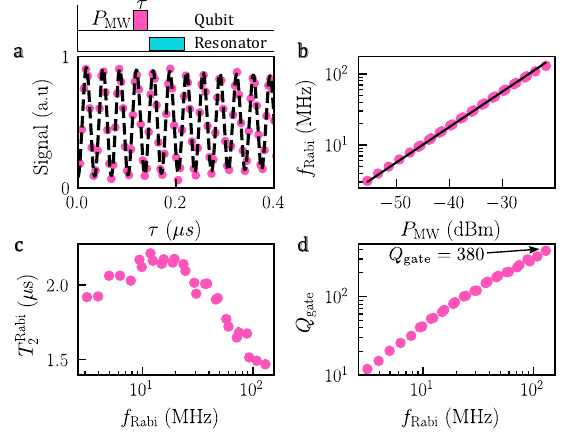}

   \caption{\textbf{Flopping mode qubit performance} (\textbf{a}) Rabi oscillations fitted to a damped sinusoidal function to extract $f_\mathrm{Rabi}$ and $T_2^\mathrm{Rabi}$ (supplementary section~\ref{sec:rabi}) with the pulse scheme sketched above. (\textbf{b}) Rabi frequency as a function of drive power applied at the chip. Black solid line is a guide to the eye with slope $0.5$. (\textbf{c}) Extracted $T_2^\mathrm{Rabi}$ as a function of $f_\mathrm{Rabi}$. (\textbf{d}) Gate quality factor $Q_\mathrm{gate}$ as a function of $f_\mathrm{Rabi}$.}
    \label{fig2}
\end{figure}
\section*{Time-domain measurements}

To assess the single qubit performances, we set the magnetic field to \SI{227}{mT} leading to  $f_\mathrm{qubit}=$~\SI{4.5}{GHz} at the charge sweet spot $\epsilon=0$ (see also Fig.~\ref{fig1}~(\textbf{d})) in a configuration mitigating decoherence, as analyzed later in this paper in Fig.~\ref{fig4}~(\textbf{c},~\textbf{d}).
Coherent control of the FM quantum state is achieved by applying resonant microwave (MW) pulses to one gate of the DQD followed by a MW pulse at $f_\mathrm{r}$ applied to the readout resonator to infer the qubit state \cite{2021_Blais}, see supplementary section~\ref{sec:dataaqui} for details.
Fig.~\ref{fig2}~(\textbf{a}) shows a representative pulsed dispersive readout signal depending on the MW burst time $\tau$, revealing Rabi oscillations of the driven FM qubit. We extract
the Rabi frequency $f_\mathrm{Rabi}$ and characteristic decay time $T_2^\mathrm{Rabi}$ for different applied MW power, see Fig.~\ref{fig2}~(\textbf{b}) and (\textbf{c}) 
respectively. The Rabi frequency grows linearly with drive amplitude up to $\SI{130}{MHz}$ without noticeable saturation. At the same time, $T_2^\mathrm{Rabi}$ slightly increases with increasing Rabi frequency with a maximum
of $\SI{2.1}{\micro s}$ around $f_\mathrm{Rabi}=\SI{20}{MHz}$ before it decreases, see
Fig.~\ref{fig2}~(\textbf{c}). From these quantities we compute the single-gate quality factor: $Q_\mathrm{gate} = 2 \cdot f_\mathrm{Rabi}T_2^\mathrm{Rabi}$ \cite{2022_Stano}, which quantifies the number of consecutive operations that can be performed until the coherence of the qubit is lost. 
With a maximum of $380$ at the largest drive power, single qubit fidelities of up to $99.9\%$ can be expected, close to state of the art values for silicon spin qubits \cite{2022_Stano}. This marks an improvement of more than an order of magnitude over the state-of-the-art for FM spin qubits \cite{Croot, 2023_Hu}.

\begin{figure*}[htbp]
    \includegraphics[width=1\linewidth]{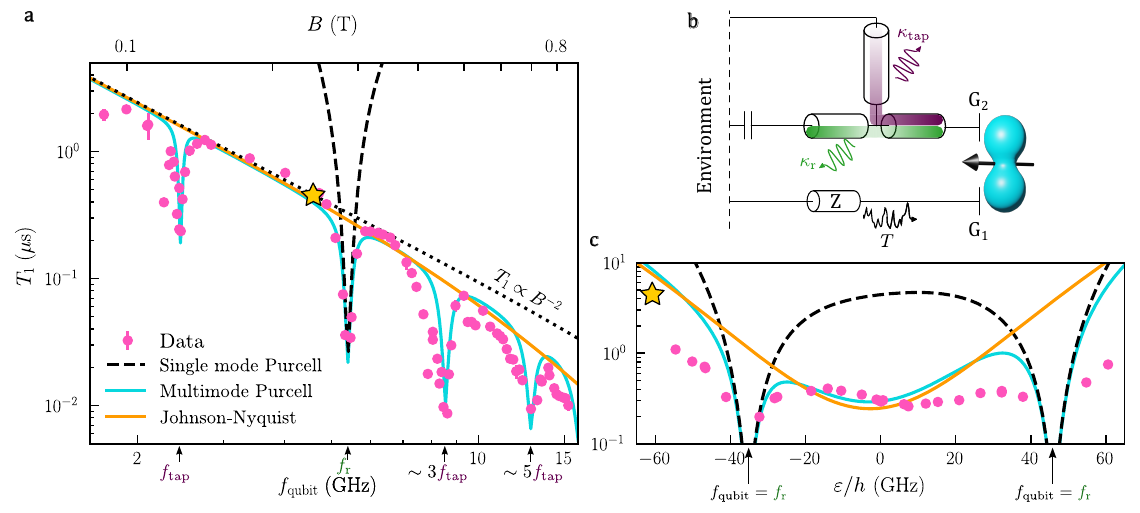}
    \caption{\textbf{Relaxation:} (\textbf{a}) Measurement of $T_1$ at the sweet-spot as a function of $f_\mathrm{qubit}$, tuned with the magnetic field. The dotted line highlights a $T_1 \propto B^{-2}$ dependence. (\textbf{b}) Representation of the qubit's microwave environment. The resonator (resp. tap) electric field shown in green (resp. violet) couples to the qubit via gate $G_2$. Gate $G_1$ is connected to a DC-line and fast line of effective impedance $Z$ at temperature $T$. (\textbf{c}) Measurement of $T_1$ as a function of $\varepsilon$ at $B=\SI{0.255}{T}$ resulting in $f_\mathrm{qubit}=\SI{4.6}{GHz}$ at $\varepsilon=0$. This working point is indicated in (\textbf{a}) by the yellow star. In (\textbf{a}) and (\textbf{c}), the dashed black line corresponds to the Purcell effect from the readout resonator fundamental mode only,  the cyan line to the multimode Purcell model and the orange line to the Johnson Nyquist relaxation.}
   \label{fig3}
\end{figure*}
Such high values for $Q_\mathrm{gate}$ raise the question on the noise sources limiting coherence. With the excellent tunability of the FM qubit discussed above, we thoroughly study the mechanisms behind spin relaxation and dephasing in the following.


\section*{{Relaxation by photon emission}}

The lifetimes of FM spin qubits reported so far lie between $\SI{100}{ns}$ \cite{2024_Dijkema} and $\SI{3}{\micro s}$ \cite{2018_Mi}, several orders of magnitude below the millisecond-lived spin qubits in single QDs \cite{BurkardReview, 2022_Stano}. This suggests that the strong spin-photon coupling comes at a cost in lifetime. Spins in Si QDs generally relax through their hybridization to the charge, which relaxes via phonons \cite{BurkardReview}. Due to the enhanced charge-character of the flopping-mode qubit, a common viewpoint is that phonon emission should limit the relaxation time \cite{Benito2019,Fang2023}.

We present in Fig.~\ref{fig3} (\textbf{a}) measurement of the relaxation time $T_1$ (see supplementary section \ref{sec:dataaqui} for details) as a function of $f_\mathrm{qubit}$ using the magnetic field as control knob. $T_1$ shows multiple  Lorentzian dips on a decreasing background for values ranging from $\SI{2}{\micro s}$ down to $\SI{10}{ns}$.

On the background, $T_1$ follows a $\sim B^{-2}$ trend which cannot be accounted for by a phonon-limited lifetime, generally scaling as $B^{-x}$ with $x\ge3$ for hole spins \cite{YM_T1, Fang2023}. Instead, we ascribe the $T_1$ dependence to the radiative decay of the qubit enhanced by Purcell effect \cite{Purcell, Houck_T1, Bertet_T1} at frequencies corresponding to the different electromagnetic modes present on the device chip. Fig.~\ref{fig3}~(\textbf{b}) depicts the microwave environment of the qubit. The DQD is coupled to the main readout resonator highlighted in green but also to a lambda/4 resonator (highlighted in purple) of fundamental frequency $f_\mathrm{tap}$, which is formed by the DC-tap used to DC-bias the gate coupled to the main resonator. In the frequency range explored in Fig.~\ref{fig3}~(\textbf{a}) we identify enhanced relaxation at the fundamental resonance of the readout resonator and at three harmonics of the tap mode at $f_\mathrm{tap}$, $\sim3f_\mathrm{tap}$ and $\sim5f_\mathrm{tap}$. Considering a single spin coupled to one electromagnetic mode, Purcell effect arises from the \textit{dressing} of the spin state with photons leaking into the environment at a rate $\kappa$. This process results in the qubit relaxing at a rate \cite{2014_Sete}:
\begin{equation}
    \frac{1}{T_1} = \frac{\kappa}{2} \left( 1- \frac{|\Delta|}{\sqrt{\Delta^2 + 4g_{\mathrm{s}}^2}}\right)
    \label{eq:purcell}
\end{equation}
with $\Delta$ the energy difference  between the spin and the mode.

\begin{figure}[htbp]
    \includegraphics[width=\columnwidth]{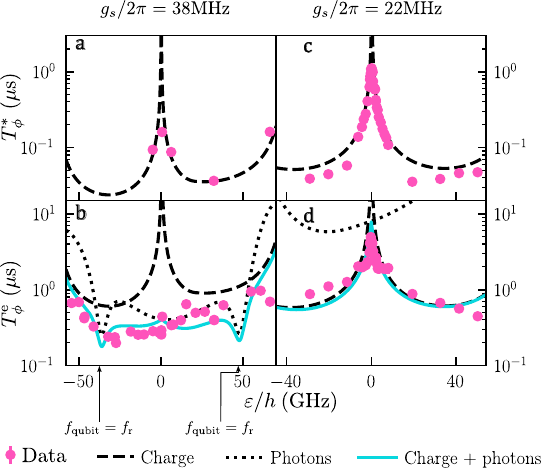}
   \caption{ \textbf{Dephasing:} (\textbf{a}) Ramsey ((\textbf{b}) Hahn echo) dephasing times as a function of $\varepsilon$, with a spin-photon coupling of \SI{38}{MHz} at the anticrossing with the resonator. These measurements were done during the first cooldown. As $f_\mathrm{qubit}$ increases with $\varepsilon$, it crosses the main resonator frequency at finite $\varepsilon$ (indicated by black arrows). (\textbf{c}) Ramsey ((\textbf{d}) Hahn echo) dephasing times as a function of $\varepsilon$ in a second cool-down with more attenuation and with a spin-photon coupling of \SI{22}{MHz}. The black dashed line corresponds to dephasing times limited by charge noise (see supplementary section \ref{sec:charge_noise}), whereas the black dotted line corresponds to photon shot noise limiting dephasing times (see supplementary section \ref{sec:photon_noise}). The cyan line combines both mechanisms (sum of the rates). (\textbf{a,~b}) and (\textbf{c,~d}) are measured with $f_\mathrm{qubit}=\SI{4.6}{GHz}$ and $\SI{4.5}{GHz}$ at $\epsilon=0$ respectively, corresponding to the point indicated with a star symbol in Fig.~\ref{fig1} and Fig.~\ref{fig3}.}
    \label{fig4}
\end{figure}
As the readout resonator and its coupling to the FM qubit are well characterized ($\kappa$ and $g_s$ known), we can compute $T_1$ near $f_\mathrm{r}$ with Eq.~\ref{eq:purcell} (black dashed line in Fig.~\ref{fig3}~(\textbf{a})), which is in quantitative agreement with the data. When other modes are present, their contributions add up, leading to a so-called multimode Purcell effect \cite{Houck_T1}. With a simplified model (see supplementary section~\ref{sec:relax_purcell}) assuming coupling to the resonator and tap harmonics up to a frequency cut-off of $\SI{200}{GHz}$, we reproduce the relaxation over the whole frequency range with good agreement, see cyan line in Fig.~\ref{fig3}~(\textbf{a}). However, we cannot exclude Johnson-Nyquist relaxation \cite{2005_Ithier,YM_T1} \textit{via} the other control lines of the device, depicted as a global impedance $Z$ in the schematic of Fig.~\ref{fig3}~(\textbf{b}). In fact, the $T_1$ background is also well reproduced (orange line in Fig.~\ref{fig3}~(\textbf{a}) by electrical noise through an impedance of $\SI{460}{\Omega}$ at a temperature of $200$\,mK (see supplementary section~\ref{sec:relax_JN}). Nevertheless, the relaxation of the FM qubit is limited by radiative decay.

For completeness, we examine the detuning dependence of $T_1$ in Fig.~\ref{fig3}~(\textbf{c}). The behavior is again well captured by our model except at large detuning where $T_1$ is overestimated as we neglect a possible spin-photon coupling when the spin is fully localized in one QD \cite{Yu}. In summary, photon emission successfully reproduces our $T_1$ measurements with remarkable agreement over the whole range of experimental parameters (magnetic field and detuning).


\section*{{Photon shot-noise dephasing}}

With energy relaxation understood, we now turn on to the study of pure dephasing. As for any electrically driven spin qubit \cite{2017_Yoneda, 2022_PiotBrun}, the common viewpoint is that the FM qubit should undergo dephasing due to charge noise \cite{2018_Mi, 2018_Samkharadze, Benito2019, Yu, Croot, 2024_Dijkema}. However, as already shown in Fig.~\ref{fig1}~(\textbf{b}) and (\textbf{d}), the FM qubit comes by construction with a first-order sweet spot, which should protect it from detuning noise \cite{2005_Ithier,Benito2019}. Consequently, we study the $\varepsilon$-dependence of the Ramsey (resp. Hahn echo) dephasing times $T_{\phi}^*$ (resp. $T_{\phi}^\mathrm{e}$), see supplementary section~\ref{sec:meas_dephasing} for details.

Setting $f_\mathrm{qubit}$ to \SI{4.5}{GHz}, Fig.~\ref{fig4}~(\textbf{a}) shows that $T_{\phi}^*$ peaks at the sweet-spot ($\varepsilon=0$), where it reaches $\SI{160}{ns}$, and drops to a minimal value of $\SI{30}{ns}$ at finite $\varepsilon$ before increasing again at even larger $\varepsilon$. This dropping behavior is well described by a linear coupling to charge-noise, where the susceptibility of the qubit to charge noise ($\propto |\frac{\partial f_\mathrm{qubit}}{\partial \varepsilon}|$) vanishes at the sweet-spot and is maximal at intermediate $\varepsilon$. The black-dashed line in Fig.~\ref{fig4}~(\textbf{a}) is the expected dephasing time assuming a 1/f detuning-like charge noise with a power spectral density $S_\varepsilon (f)=A_\varepsilon \times \SI{1}{Hz}/f$ of amplitude $\sqrt{A_\varepsilon} \sim \SI{0.2}{\micro eV /\sqrt{Hz}}$.

Strikingly, the echo dephasing time $T_{\phi}^e$, shown in Fig.~\ref{fig4}~(\textbf{b}), does not reveal any sweet-spot behavior. It has a minimal value of $\SI{260}{ns}$ around $\varepsilon=0$ and increases away from it up to $\sim \SI{1}{\micro s}$. This behavior is in disagreement with an echo dephasing time mainly limited by detuning noise as revealed by the discrepancy between the expected dephasing (black dashed line) and the experimental data. Furthermore, the ratio $T_{\phi}^\mathrm{e}/T_{\phi}^*\sim 1.5$ at $\varepsilon=0$ is unexpectedly small for dominant low-frequency charge noise as it should be efficiently filtered by the refocusing pulse of a Hahn echo sequence \cite{2005_Ithier}. These observations indicate that a high-frequency noise source is at play.

As we show below, the limiting noise source is compatible with dephasing due to thermal photon-number fluctuations (shot noise) in the different resonators coupled to the FM qubit, which shift its frequency \textit{via} the a.c. Stark effect. In the dispersive regime, the dephasing rate due to thermal photon-number fluctuations in a resonator can be written as \cite{2007_Clerk}:
\begin{equation}
   \frac{1}{T_{\phi}^{th}} = \chi^2\frac{\bar{n}(\bar{n}+1)}{\kappa}
  \label{T2_photon_noise^th}
\end{equation}
with $\chi= 2g_\mathrm{s}^2/\Delta$ the dispersive shift of the qubit frequency per photon and $\bar{n}$ the resonator average thermal photon population. 
Taking into account the two closest modes to the FM qubit, namely the readout resonator at \SI{5.4}{GHz} and the DC-tap mode at \SI{2.4}{GHz}, we reproduce the echo dephasing times around the sweet-spot with photon-number fluctuations assuming photonic temperatures of $\SI{80}{mK}$ for the readout-resonator and $\SI{230}{mK}$ for the tap mode respectively, see dotted line in Fig.~\ref{fig4}~(\textbf{b}) and supplementary section \ref{sec:photon_noise}. Such photonic temperatures are high compared to the $\SI{50}{mK}$ regularly achieved in the cQED community \cite{2018_Yan, Devoret_attenuator}, but not un-realistic given the attenuation scheme and absence of careful microwave hygiene in typical spin qubit setups, (see supplementary section \ref{sec:setup}). For $|\varepsilon| > 2 t_\mathrm{c}$, the photon induced dephasing is largely reduced as $\chi$ decreases when the spin gets localized in one QD and $T_\phi^e$ is then mainly limited by charge noise. Eventually, combining the dephasing rate originating from thermal photon shot noise and detuning charge noise we are able to capture the detuning behavior of $T_{\phi}^e$, see cyan line in Fig.~\ref{fig4}~(\textbf{b}).

To complement the experimental findings of Fig.~\ref{fig4}~(a,b) we thermally cycle the device with a modified cryogenic setup with increased attenuation on the high frequency control lines to reach possibly colder photonic temperatures. At the same time, the spin-photon coupling, at resonance with the main resonator, was found to be reduced to \SI{22}{MHz} compared to the initial \SI{38}{MHz}. The close to two-fold reduction of $g_s$ should directly lead to a roughly ten-fold increase in $T_\phi^{th}$ given the fast scaling $\propto g_s^4$  expected from Eq.~\ref{T2_photon_noise^th}. In Fig.~\ref{fig4}~(\textbf{c}) and (\textbf{d}) we report the Ramsey and echo dephasing times for this new cool-down. At $\varepsilon=0$, we observe a clear sweet-spot behavior for both $T_\phi^*$ and $T_\phi^e$ associated with a ten-fold increase compared to Fig.~\ref{fig4}~(\textbf{a}) and (\textbf{b}). The maximum value of $T_\phi^*=$\SI{1.2}{\micro s} at the sweet-spot is now potentially limited by hyperfine interaction \cite{2022_PiotBrun}. Concerning the \SI{5}{\micro s} observed for $T_\phi^e$, it might still be limited by photonic noise as suggested by the observation of a pure exponential decay (see supplementary section \ref{sec:betas}). Assuming only photonic noise as the remaining dephasing mechanism at $\varepsilon=0$ for $T_\phi^e$, we find photonic temperatures very similar to the ones found before adding additional attenuators as the increase in $T_\phi^e$ can entirely be captured by the change of $g_\mathrm{s}$, see dotted line in Fig.~\ref{fig4}~(\textbf{d}). Finding similar effective modeling temperatures for both cool-downs suggests that the eleveated photonic temperatures are not due to the connection to poorly thermalized high frequency control-lines but rather result from stray radiation in the cryostat, hence calling for rigorous improvements in the experimental setup \cite{2011_Barends, 2011_Corcoles}.

\section*{Discussion and Outlook}
Besides the ability of FM spin qubits to strongly interact with microwave photons in superconducting resonators \cite{2018_Mi, 2018_Samkharadze, Yu}, we establish here, for the first time, that a FM spin qubit can also be a viable single spin qubit with a demonstrated single-qubit gate quality factor of 380. While ultra-fast spin qubits (Rabi frequency $\sim 100$\,MHz) generally achieve poor gate fidelities \cite{Watzinger_Ncomms_2018, 2021_Froning, 2022_Wang}, we show here that even with gate times of a few ns, the built-in noise-insensitive detuning point of the FM qubit allows to reach quality factors competing with that of state of the art spin qubits. Importantly, our study pinpoints the microwave environment to be the major source of decoherence and not the semiconducting environment (phonon and charge noise) as typically experienced by QD-based spin qubits \cite{2022_Stano}. Coherence limitations are compatible with radiative decay and dephasing due to thermal photon shot-noise, two mechanisms well-known from cQED. Hence, mitigation strategies inherited from the superconducting qubit community can be readily applied to further enhance the FM qubit performances. 

For thermal photon shot noise,  adequate filtering and shielding can reduce the photonic temperature down to \SI{60}{mK} or below \cite{2018_Yan, Devoret_attenuator} allowing potentially a 20-fold increase in dephasing time. In that case, hyperfine interaction or charge noise \cite{2022_PiotBrun, Hendrickx2024} will likely become the limiting factors. The hyperfine dephasing can be minimized with isotopic purification of silicon \cite{BurkardReview} and with the level of detuning noise ($\sim \SI{0.2} {\micro eV/\sqrt{\mathrm{Hz}}}$) inferred from this study, we estimate dephasing times above $\SI{100}{\micro s}$ at the sweet-spot for second-order coupling \cite{2005_Ithier}, in-line with the best value reported for electrically driven spin qubits~\cite{2022_Stano}.


Concerning energy relaxation, radiative decay through Johnson Nyquist or Purcell effect can be mitigated by careful microwave engineering including filtering, higher quality superconducting cavities and Purcell filters \cite{2021_Blais}. At this stage, spin cQED engineering is at its beginning, and the exact gain on the resonator quality factor and photonic temperature is difficult to evaluate, motivating future experiments. Moreover, relaxation through phonon emission, which is likely the ultimate limit for the lifetime, was not detected in our study and remains to be evaluated and observed.

With further improvements in hybrid spin cQED architectures, we envision that FM spin qubits will play a vital role in creating large quantum systems for quantum information processing and simulation intermixing microwave photons and semiconductor quantum dot-based spin qubits.




\clearpage

\textbf{Author contributions:}

C.Y. fabricated the NbN-circuitry with the help from S.Z.. L.N. performed the measurements with the help of S.Z. L.N. analyzed the data with inputs from R.M., E.D. and S.Z. J.C.A.U. developed the theoretical model and helped in the interpretation of the data. L.N., R.M.and S.Z. co-wrote the manuscript with inputs from all authors. H.N, and B.B. were responsible for the front-end fabrication of the device. S.Z. supervised the work.

\begin{acknowledgments}
We thank J.-L. Thomassin and F. Gustavo for help in the fabrication of the NbN circuitry and M. Boujard and I. Matei for technical support in the lab. Silvano De Franceschi is acknowledged for fruitful discussions and careful proofreading of the manuscript. We are grateful to Yann-Michel Niquet for insightful discussions and we also thank Clemens Winkelmann for proofreading of the manuscript.

This research has been supported by the European Union’s Horizon 2020 research and innovation programme under grant agreements No. 951852 (QLSI project), No. 810504 (ERC project QuCube) and No. 759388 (ERC project LONGSPIN), No. 101174557 (QLSI2) and by the National strategy France 2030 under the project PEPR PRESQUILE - ANR-22-PETQ-0002 and PEPR MiraclQ ANR-23-PETQ-0003. S. Zihlmann acknowledges support by the spin-photon PEPR chair. J. C. A. U. is supported by the Spanish Ministry of Science, innovation, and
Universities through Grants PID2023-
148257NA-I00 and RYC2022-037527-I.

\end{acknowledgments}

\textbf{Data availability}
The datasets generated and analyzed during this current study are available at https://doi.org/10.5281/zenodo.18683389.


\onecolumngrid
\newpage
\setcounter{figure}{0}
\section*{Supplementary information}
\renewcommand\thesection{\arabic{section}} 
\def\theequation{S\arabic{equation}}
\renewcommand{\thetable}{S\Roman{table}}
\renewcommand{\thefigure}{S\arabic{figure}}
\resumetoc


\author{Léo Noirot}
\affiliation{Univ. Grenoble Alpes, CEA, Grenoble INP, IRIG-Pheliqs, Grenoble, France.}
\author{Cécile X. Yu}
\affiliation{Univ. Grenoble Alpes, CEA, Grenoble INP, IRIG-Pheliqs, Grenoble, France.}
\affiliation{QuTech and Kavli Institude of Nanoscience, Delft University of Technology, P.O. Box 2046, Delft, 2600 GA, Delft, The Netherlands.}
\author{José C. Abadillo-Uriel}
\affiliation{Univ. Grenoble Alpes, CEA, IRIG-MEM-L\_Sim, Grenoble, France.}
\affiliation{Instituto de Ciencia de Materiales de Madrid, Consejo Superior de Investigaciones Cientificas, Madrid 28049, Spain.}
\author{Étienne Dumur}
\affiliation{Univ. Grenoble Alpes, CEA, Grenoble INP, IRIG-Pheliqs, Grenoble, France.}
\author{Heimanu Niebojewski}
\affiliation{Univ. Grenoble Alpes, CEA, LETI, Minatec Campus, Grenoble, France.}
\author{Benoit Bertrand}
\affiliation{Univ. Grenoble Alpes, CEA, LETI, Minatec Campus, Grenoble, France.}
\author{Romain Maurand}
\affiliation{Univ. Grenoble Alpes, CEA, Grenoble INP, IRIG-Pheliqs, Grenoble, France.}
\email{romain.maurand@cea.fr}
\author{Simon Zihlmann}
\affiliation{Univ. Grenoble Alpes, CEA, Grenoble INP, IRIG-Pheliqs, Grenoble, France.}


\tableofcontents

\section{Setup and flopping mode parameters}\label{sec:setup}

\subsection{Device and measurement setup} \label{sec:setup}
\begin{figure*}[htbp!]
    \includegraphics[height=18.5cm, keepaspectratio]{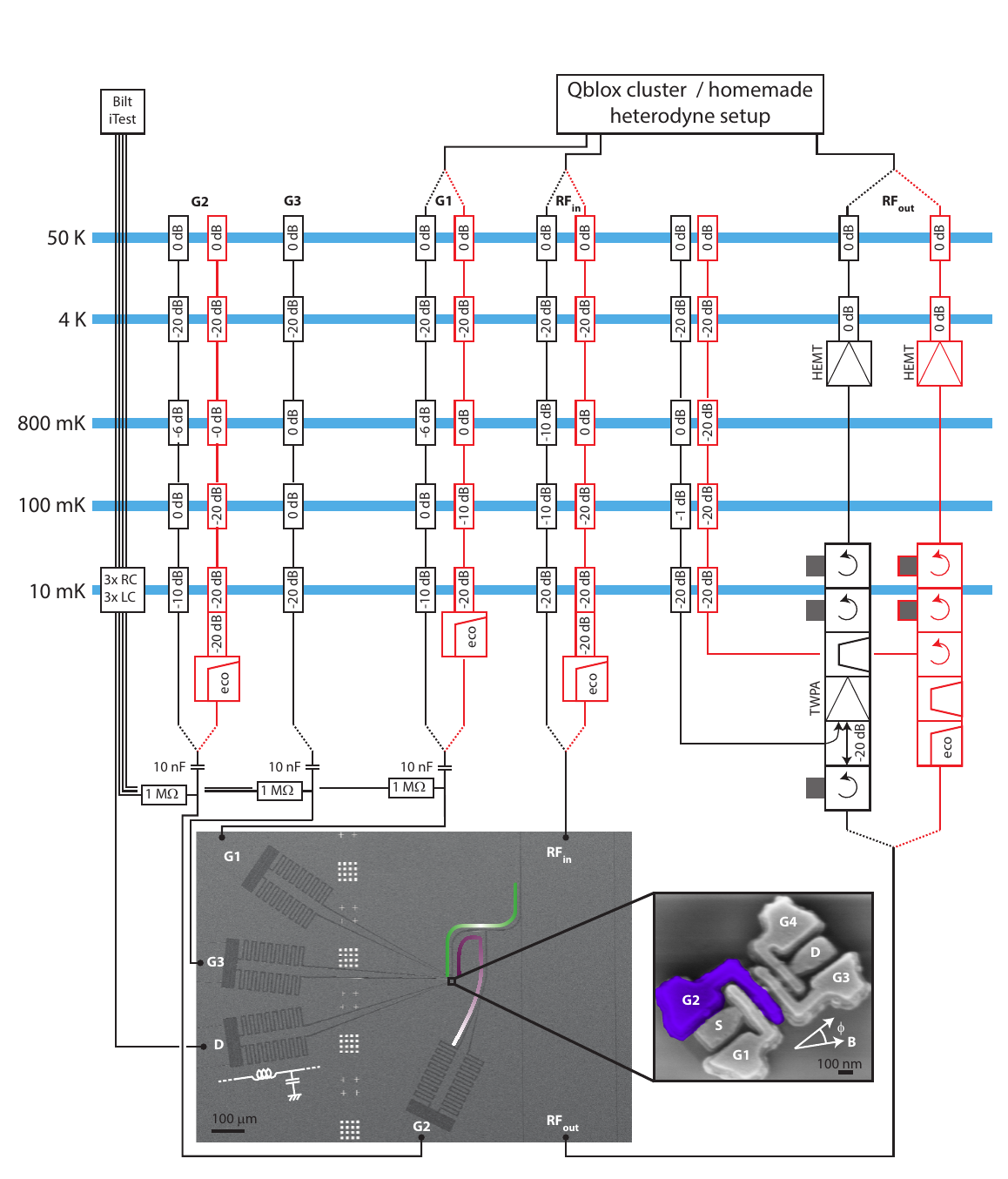}

   \caption{\textbf{Measurement setup and device schematic:} Detailed wiring of the measurement setup. Black corresponds to the wiring of the setup in the first cool-down, whereas red indicates the changes made to the setup for the second cool-down. An overview scanning electron micrograph of a nominally identical device is shown, with a zoom-in of the silicon nanowire transistor. The magnetic field is applied in-plane of the sample with an angle $\phi$ with respect to the silicon nanowire axis. Source (S) of the device is hard grounded to the NbN ground plane and G1 and G4 are shorted together at the device level. For simplicity, the gate line is called G1, see Ref.~\cite{Yu} for more details. The zero-point voltage fluctuation of the two fundamental modes of the readout resonator as well as the DC-tap are sketched as green to white and purple to white gradients respectively. The white LC-schematics indicates the on-chip low pass filters consisting of a nanowire inductor (L) and a finger capacitor to ground (C).
   }
    \label{device}
\end{figure*}

The device and its fabrication used in this study are described in detail in Ref.~\cite{Yu}. In short, it consists of a natural silicon on insulator nanowire transistor with four overlapping gates in series, see Fig.~\ref{device}. Gate G2 is galvanically connected to a voltage anti-node of a microwave resonator patterned in a niobium nitride (NbN) film. All DC connections are fitted with LC low-pass filters, see Fig.~\ref{device}.

All measurements are performed in a dilution refrigerator equipped with a three axes vector magnet at a base temperature of \SI{8}{mK}. A detailed wiring schematic is presented in Fig.~\ref{device}. The DC gate voltages are supplied by a BE2231 card in a Bilt rack from Itest
and are low pass filtered at mixing chamber temperature (multi stage LC and RC filters). Microwave transmission measurements (RF$_\mathrm{in}$ to RF$_\mathrm{out}$) are performed either with a vectorial network analyser (VNA) M5180 from Copper Mountain, a Qblox cluster or with a homemade hetereodyne setup (consisitng of a Zurich UHF, Holzworth RF synthesizer and a Tektronix AWG). Microwave excitations are either applied to RF$_\mathrm{in}$ or to G1/G2 through their bias-Tees.

\subsection{Modes of the tap and readout resonator}

\begin{figure*}[htbp]
    \includegraphics[width=1\linewidth]{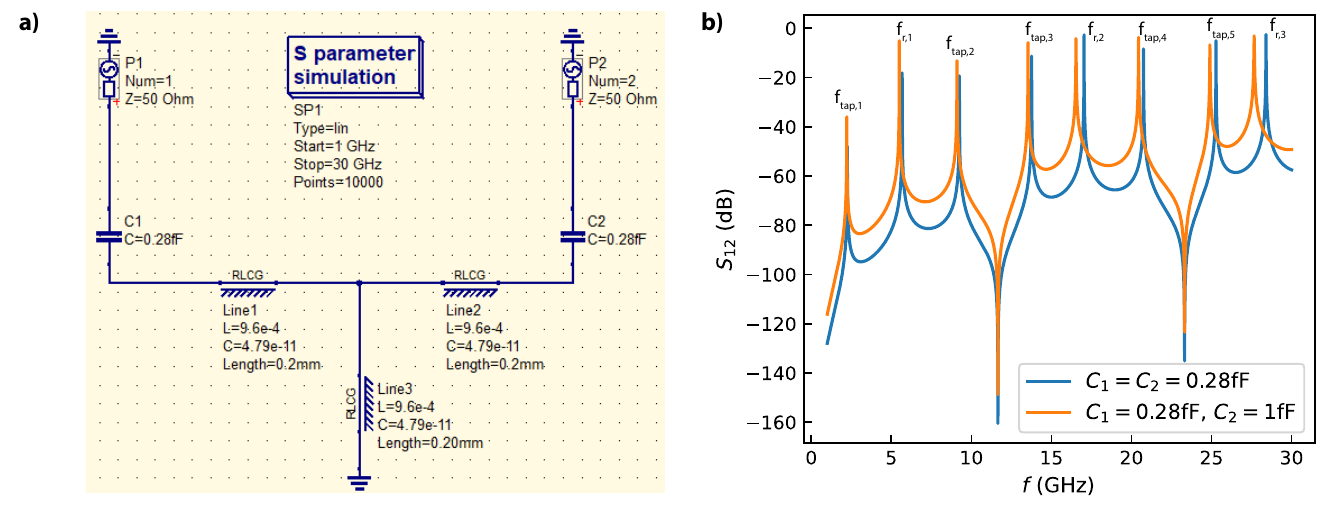}

   \caption{\textbf{Numerical model of the resonator circuit.} (\textbf{a}) The software QUCS is used to simulate an equivalent circuit to illustrate the mode spectrum of the resonator with its DC-tap. RLCG elements represent waveguides with inductance (L) and capacitance (C) per unit length in SI units. C1 and C2 model the capacitive loading of the ends of the resonator by the QD device on one side and by the transmisison line on the other side. For simplicity, the LC-filter at the DC tap is replaced with a short to ground. (\textbf{b}) Simulated transmission spectrum showing the mode spectrum for a symmetric loading of both resonator arms ($C1=C2$) and for an asymmetric loading ($C_1\neq C_2$)), which leads to distortions of the otherwise symmetric spectrum.
   }
    \label{resonator_model}
\end{figure*}

The different resonating modes of the NbN circuitry as schematically outlined in Fig.~\ref{fig3}~(\textbf{b}) of the main text as well as in Fig.~\ref{device} are discussed in the following. The resonator with its DC-tap consists of three equal segments of a coplanar waveguide in a T-shape geometry; see Fig.~\ref{device} for an image and Fig.~\ref{resonator_model} for an equivalent electrical circuit. Similar resonator designs, which allow for DC biasing of the central conductor of the resonator, have been previously reported in the literature~\cite{CollardFiltering, 2025_Koolstra}. Fig.~\ref{resonator_model} shows an electrical model and corresponding resonance modes. For specificity, we refer to them in the supplementary information with the indecis $m,n$ where $m=\mathrm{r/tap}$ indicates the medium of the resonance (r for the readout resonator and tap for the tap) and $n$ refers to the corresponding number of harmonic $n$, with $n=1$ the fundamental mode.

\begin{figure*}[htbp]
    \includegraphics[width=1\linewidth]{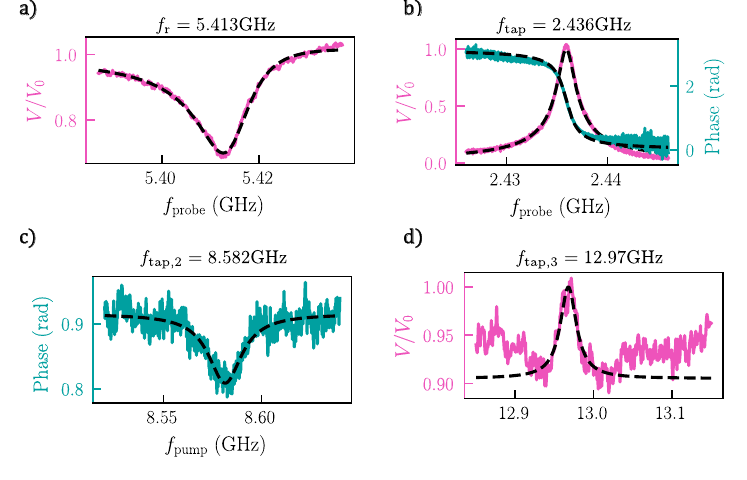}

   \caption{\textbf{Readout resonator and dc tap harmonics} (\textbf{a}) Transmission in amplitude through the feedline as a function of VNA frequency, with less than one photon in the cavity.(\textbf{b}) Transmission through the tap, taking the feedline input as input and the tap end as output, as a function of VNA frequency. The power applied corresponds to $\SI{-46}{dBm}$ at the chip assuming a cable losse of $\SI{6}{dB}$. (\textbf{c}  (resp. \textbf{d})) Phase  (resp. normalized amplitude) of the signal transmitted through the feedline at $f_\mathrm{probe}=\SI{5.413}{GHz}$ as a function of pump tone frequency. The power applied at the chip, neglecting cable losses, corresponds to $\SI{-95}{dBm}$  (resp.$\SI{-73}{dBm}$). All measurements are carried out at $B=0$ and with the charge qubit completely detuned from the cavity (e.g. large $\varepsilon$).
   }
    \label{modes}
\end{figure*}

Fig.~\ref{modes}~(\textbf{a}) shows the transmission (from RF$_\mathrm{in}$ in to RF$_\mathrm{out}$), measured with a VNA, as a function of probe frequency $f_\mathrm{probe}$. At the resonance frequency of the fundamental mode of the readout resonator, the transmission exhibits a dip in amplitude  which we fit to extract $f_\mathrm{r,1} = \SI{5.413}{GHz}$ and the resonator internal (resp. coupling) quality factor $Q_\mathrm{r,1}^i = 527$ (resp. $Q_\mathrm{r,1}^c = 1190$), corresponding to the photonic decay rate $\kappa_\mathrm{r,1}^i/2\pi = \SI{10.2}{MHz}$ (resp. $\kappa_\mathrm{r,1}^c/2\pi = \SI{4.5}{MHz}$). The total mode losses are $\kappa_\mathrm{r,1}/2\pi=\kappa_{r,1}^i/2\pi+\kappa_\mathrm{r,1}^c/2\pi=\SI{14.8}{MHz}$.

Fig.~\ref{modes}~(\textbf{b}) shows the transmission as a function of $f_\mathrm{probe}$ through the tap (from RF$_\mathrm{in}$ to G2). The data exhibits a characteristic response of a resonator probed in transmission, close to the frequency expected for the tap's fundamental mode: $f_\mathrm{r,1}/2\sim \SI{2.7}{GHz}$ \cite{CollardFiltering}. We extract from it $f_\mathrm{tap,1}=\SI{2.436}{GHz}$ and a resonance linewidth of $\SI{1.7}{MHz}$. As this measurement is done at high power (in the multi-photon regime), which reduces the dielectric losses \cite{2018_Shearrow}, the observed linewidth is a lower bound to the loss rate $\kappa_\mathrm{tap,1}$.


In order to characterize the higher harmonic modes, we perform two-tone spectroscopy, where the transmisison through the feedline at $f_\mathrm{r,1}$ is measured while another pump tone is applied at RF$_\mathrm{in}$, see Fig.~\ref{modes}~(\textbf{c}) and (\textbf{d}). 
Varying the pump frequency $f_\mathrm{pump}$ across a mode resonance shifts the resonator frequency as expected from Cross-Kerr effect \cite{2021_Yu}, leading to a change in transmission. We extract from it the resonant frequencies of the second and third harmonics of the tap: $f_\mathrm{tap,2}=\SI{8.582}{GHz}$ and $f_\mathrm{tap,3}=\SI{12.97}{GHz}$, which are close to the expected frequencies given by $(2n-1)\cdot f_\mathrm{tap,1}$. The deviation of the observed resonant frequencies from the expected ones can be explained by the capacitive loading of the different ends that are neglected in a simple estimation where three equally long segments of transmission lines are connected in a T-shape \cite{CollardFiltering}. In addition, these measurements give also insights into the linewidths, which we extract to be $\SI{12.9(5)}{MHz}$ and $\SI{15.7(9)}{MHz}$ respectively. Again. these measurements are done at high pump power thus underestimating the bare linewidths of these resonances.

\subsection{Hole spin flopping mode model}\label{sec:FM model}



To describe our system, we use the hole spin flopping mode Hamiltonian introduced in 
Ref.\cite{Yu}, which can be written as:
\begin{equation}
    H_{DD} = -\frac{\varepsilon}{2}\tau_z + \mu_B \frac{\tau_L g_L^* + \tau_R g_R^*}{2}B \sigma_z + t_{\uparrow \uparrow}\tau_x - t_{\uparrow \downarrow} \tau_y \sigma_y.
    \label{hamiltonian}
\end{equation}

Here, the basis set is $\{ |L, \uparrow \rangle, |L, \downarrow \rangle, |R, \uparrow \rangle, |R, \downarrow \rangle \}$ where $\{ |L, \uparrow \rangle, |L, \downarrow \rangle\}$ are the Zeeman-split states of the left dot and $\{ |R, \uparrow \rangle, |R, \downarrow \rangle\}$ are the Zeeman-split states of the right dot. $\tau_{\alpha} = |\alpha \rangle \langle \alpha |$ with $\alpha = L, R$ are the Pauli operators in position space and  $\sigma_{y, z}$ are the Pauli operators in spin space.

The first term corresponds to the bare detuning energy between both dots. The second is an effective Zeeman interaction with site-specific g-factors. The two last terms correspond to the tunneling energy renormalized by spin-orbit interaction, giving rise to a spin-conserving tunneling $t_{\uparrow \uparrow}$ and to a spin-flip tunneling $t_{\uparrow \downarrow}$ which preserve tunneling of the charge qubit at zero magnetic field: $t_{\uparrow \uparrow}^2 + t_{\uparrow \downarrow}^2 = t_c^2$. $t_{\uparrow \downarrow}$ captures both the Rashba interaction, and spin-flip mechanism owning to differences in the anisotropic Zeeman responses of the left and right single dots, usually referred to as g-matrices \cite{2018_Crippa}.

The preservation of tunneling allows-us to introduce an angle $\theta$ such that:

\begin{equation}
    t_{\uparrow \uparrow} = t_c \cos \theta \text{ and } t_{\uparrow \downarrow} = t_c \sin \theta
    \label{theta}
\end{equation}

This angle quantifies the spin-charge mixing, as $\sin ^2 \theta $ is the probability of flipping the spin while tunneling between L and R. This model neglects the electrostatic dependence of the g-matrices, which may introduce an intrinsic detuning dependence of the angle $\theta$~\cite{Yu}.

Tunnel coupling $t_c$, site-specific g factor $g_R^*$ and $g_L^*$ can be accessed experimentally; see the following sections. Eventually, the only parameter that cannot be directly measured is $\theta$. It thus can be estimated by fitting the model to the dataset of the flopping-mode qubit frequency as a function of magnetic field, see section~\ref{FM parapmeters}.





\subsection{Charge configuration}

\begin{figure*}[htbp]
    \includegraphics[width=1\linewidth]{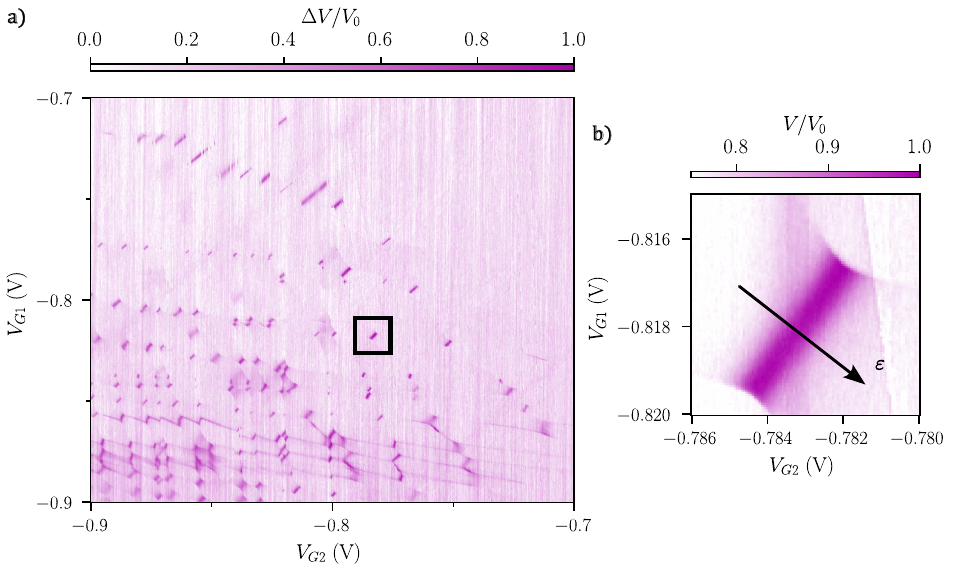}

   \caption{\textbf{Stability diagram} (\textbf{a}) Amplitude of the transmission at $f_r$, as a function of $V_{G1}$ and $V_{G2}$. A background is removed for each vertical cut to account for slow variation of the signal, likely due to spurious dots and or local changes in the electric environment to which the cavity is sensitive. Generally, these features do not interact with any charge interdot transition defined by G1 and G2. The interdot used in our study is higlighted by a black box. (\textbf{b}) Zoom on the interdot, where the black arrow defines the orientation of the $\varepsilon$-axis.}
    \label{stability}
\end{figure*}

Fig.~\ref{stability}~(\textbf{a}) shows the charge stability diagram with respect to the voltages on gates G1 and G2, with $V_{S,D}=0$ and G4 shorted to G1 at the device level. $V_{G3}$ is set to a voltage such that no charges are accumulated below G3. Localized diagonal features correspond to two orbital states being resonant, allowing a charge to be delocalized between two dots. Charges with sufficiently large coupling to the cavity will dispersively shift its frequency, resulting in a change in transmission through the feedline. The interdot studied in here is highlighted by a black box.

Fig.~\ref{stability}~(\textbf{b}) shows a zoom on the interdot, where the $\varepsilon$-axis is defined perpendicularly to the interdot charge transition.

\subsection{Estimating the lever arm}

In our previous work, the lever-arm $\alpha$, characterizing the coupling of the voltage applied on the electrodes to the  electro-chemical potential of the dots, was measured using the temperature dependence of the dispersive shift of the charge-qubit, which depends on its population \cite{Yu}. At our first cooldown, the setup was equiped with a TWPA based on aluminum Josephson-junctions from Silentwaves (Argos). The transmission through the TWPA collapes around $\SI{0.7}{K}$, preventing us from using the same method. We therefore used Landau-Zener-Majorana-St\"{u}ckleberg interferometry to perform the spectroscopy of the charge qubit, thereby measuring its frequency $f_c$ \cite{2017_Penfold-Fitch, LZMS}. The dependence of $f_c$ on voltage detuning: $\epsilon_v = \beta_1 V_\mathrm{G_1}-\beta_2 V_\mathrm{G_2}$ with $\beta_1=0.76$ and $\beta_2=0.65$ (see black arrow in Fig.~\ref{stability}~(\textbf{b})) allows us to estimate $\alpha$. For the second cooldown, we removed the TWPA and measured $\alpha$ using the temperature dependence of the charge shift (not provided here).  

\begin{figure*}[htbp]
    \includegraphics[width=1\linewidth]{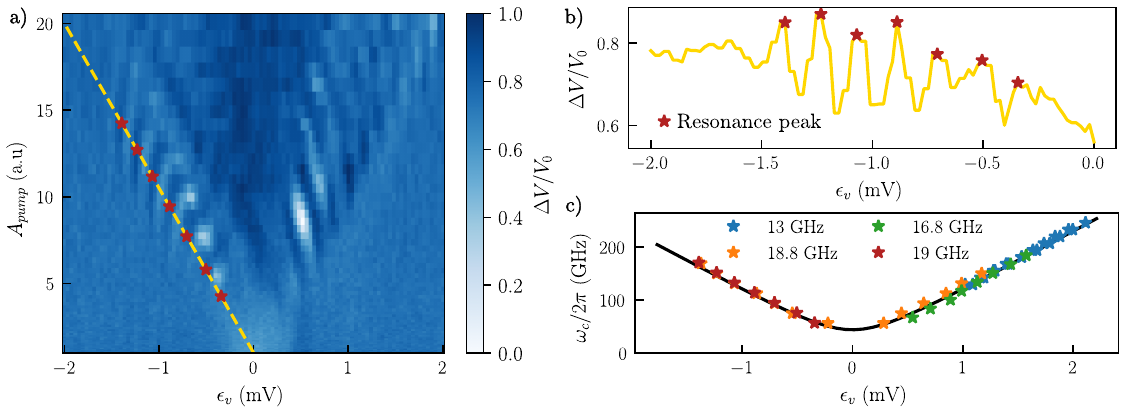}

   \caption{\textbf{Lever-arm and tunneling estimation} (\textbf{a}) Transmission amplitude 
 measured at the resonator frequency as a function of  $\varepsilon_v$ while a pump tone at frequency $f_{pump} = \SI{19}{GHz}$ of amplitude $A_\mathrm{pump}$ (in linear scale)  applied on G2.  The probe frequency is adjusted to match with the $\varepsilon_v$-dependence of the resonator frequency, and a background is removed for each vertical cut. The amplitude along the yellow-dashed is plotted in (\textbf{b}), where multiphoton peaks are represented by red stars, from which we extract $f_c$ at a given $\varepsilon_v$. We repeated the process for $f_{pump} = \SI{13}{GHz}$, $\SI{16.8}{GHz}$, $\SI{18.8}{GHz}$ and $\SI{19}{GHz}$ where the deduced values of $f_c$ are represented by stars in (\textbf{c}). The black line corresponds to Eq. \ref{cq_nrj} with $t_c = \SI{23}{GHz}$ and $\alpha = 0.47$.}
    \label{LZMS}
\end{figure*}

Fig.~\ref{LZMS} shows transmission through the feedline as function of $\epsilon_v$, while an other pump tone is applied on gate $\mathrm{G_2}$ at $f_\mathrm{pump}=\SI{19}{GHz}$. Varying the amplitude of the pump tone $A_{pump}$ reveals an interference pattern where vertical lines correspond to multiphoton excitation of the charge qubit: $f_c (\epsilon_v)=nf_\mathrm{pump}$, with $n$ the number of photons.

By analyzing cuts along the side of the pattern, we retrieve the multiphoton resonances as depicted on Fig.~\ref{LZMS}~(\textbf{b}), which allows us to reconstruct the $\epsilon_v$-dependence of $f_c$ as shown in Fig.~\ref{LZMS}~(\textbf{c}).

Repeating this measurement for different pump frequencies allowed us to consistently retrieve the charge qubit energy for each value of $f_\mathrm{pump}$. A fit of the charge qubit energy using:

\begin{equation}
    f_c = \frac{1}{h}\sqrt{(2t_c)^2 + (\alpha e \epsilon_v)^2}
    \label{cq_nrj}
\end{equation}
renders $\alpha=0.47(1)$ which we use in the following to define the detuning axis: $\varepsilon = \alpha \varepsilon_v$. It additionnaly gives a first estimation for the tunnel coupling: $t_c/h=\SI{23(1)}{GHz}$.


\subsection{Characterizing the charge qubit}\label{chargeq}
We evaluate the tunnel coupling $t_c$ and the charge-photon coupling $g_c$ by analyzing the dispersive shift of the cavity frequency caused by its interaction with the charge qubit, given by $\tilde f_\mathrm{r, 1} = f_\mathrm{r, 1} + \chi_c/2\pi$, where 

\begin{equation}
    \chi_c = \frac{g_c^2 d^2_{c}}{2\pi}\left( \frac{1}{|f_c - f_\mathrm{r, 1}|} + \frac{1}{f_c + f_\mathrm{r, 1}}\right)
    \label{chic}
\end{equation}
is the dispersive shift with $d_{c}$ the electric dipole of the charge qubit:

\begin{equation}
    d_{c} = \frac{2t_c}{\sqrt{\varepsilon^2 + (2t_c)^2}}
    \label{d01}
\end{equation}

\begin{figure*}[htbp]
    \includegraphics[width=1\linewidth]{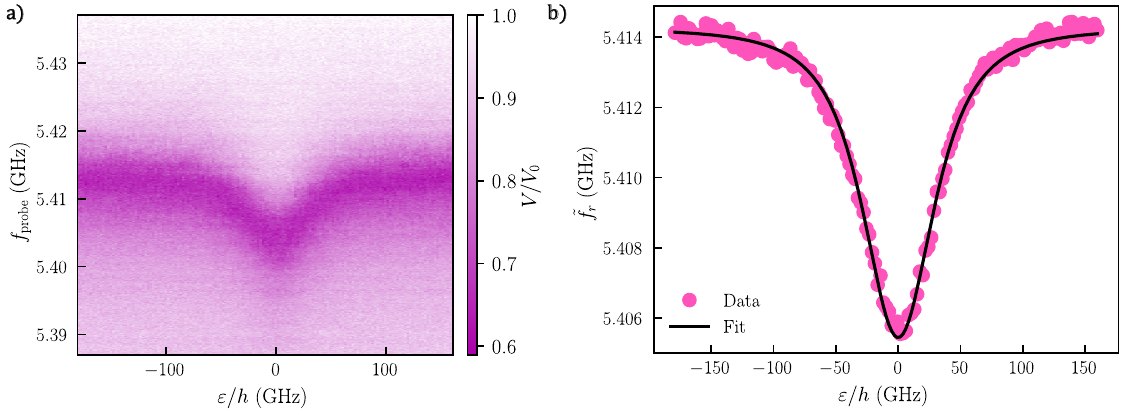}

   \caption{\textbf{Charge qubit - resonator interaction} (\textbf{a}) Normalized amplitude transmitted through the feedline as a function of probe frequency and $\varepsilon$. From each vertical line, we extract$\tilde f_\mathrm{r, 1}$ at a given $\varepsilon$ as shown in (\textbf{b}). Fitting the $\varepsilon$-dependence of $\tilde f_\mathrm{r, 1}$ with Eq. \ref{chic} yields $g_c/2\pi = \SI{437 \pm 10}{MHz}$ and $t_c/h = \SI{22 \pm 1}{GHz}$.}
    \label{chic_meas}
\end{figure*}

In Fig.~\ref{chic_meas}~(\textbf{a}), we measure the resonance of the resonator as a function of $\varepsilon$, from which we extract the $\varepsilon$-dependence of $\tilde f_\mathrm{r, 1}$ (see Fig.~\ref{chic_meas}~(\textbf{b})). Fitting the data with Eq.~\ref{chic} reproduces the dispersive shift with accuracy, and results in $g_c/2\pi = \SI{437 \pm 10}{MHz}$ and $t_c/h = \SI{22 \pm 1}{GHz}$, in excellent agreement with the previous measurement regarding the tunnel coupling. 

\subsection{g-factor of the single dots}\label{gfactors}
\begin{figure*}[htbp]
    \includegraphics{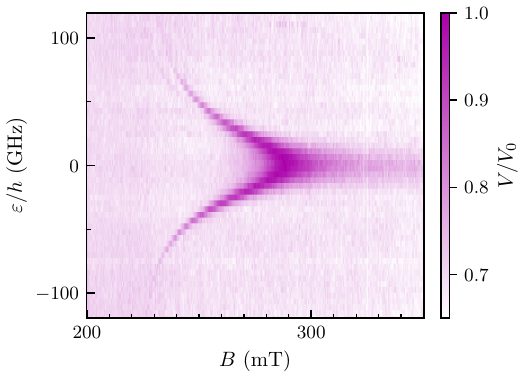}

   \caption{\textbf{Extracting the single dots' g-factors}: Normalized amplitude transmitted through the feedline at frequency $f_\mathrm{{probe}}(\varepsilon) = \tilde f_{\mathrm{r}, 1}(\varepsilon)$, as a function of $\varepsilon$ and $B$. For each detuning value, the probe frequency $f_\mathrm{{probe}}$ is adjusted to account for the $\varepsilon$-dependence of the resonator (see also Fig. \ref{chic} a)). Darker region indicates the points at which the qubit crosses the resonator: $f_\mathrm{qubit} = \tilde f_{\mathrm{r}, 1}$. The curved profile is a direct consequence of the spin-orbit-induced renormalization of the g-factor at zero detuning, and this measurement can be seen as a dual of a 2-tone spectroscopy of the sweet-spot, where increasing $B$ corresponds to reducing the drive frequency (See Fig.~\ref{fig1}~c). At $|\varepsilon|/h = \SI{80}{GHz}$, the hole is almost fully localized in a single dot ($\varepsilon /ht_c \sim 4$) resulting in $f_\mathrm{qubit} \simeq g_{L,R}^* \mu_B B/h$.}
    \label{tuning_spin}
\end{figure*}

To access the left and right quantum dot g-factors, we measure the response of the readout resonator as a function of $\varepsilon$ and magnetic field amplitude $B$ as shown in Fig.~\ref{tuning_spin}~(\textbf{a}). For a given value of $\varepsilon$, sweeping $B$ increases the spin transition energy and leads to a change in transmission when the spin is resonant with the cavity. For different values of $\varepsilon$, this measurement reconstructs the values of $B$ at which $f_\mathrm{qubit} = f_\mathrm{r, 1}$. For $|\varepsilon|/h\gg t_c$, the spin gets isolated in a single dot: $f_\mathrm{qubit} = g_{L,R}^* \mu_B B/h$ and the resonance condition becomes $g_{L,R}^* \mu_B B = hf_\mathrm{r, 1}$.

Taking the values at $\varepsilon/h = \SI{80}{GHz}$, we measure $g_L^* = \SI{1.62 \pm 0.02}{}$ and $g_R^*=\SI{1.67\pm 0.02}{}$ for a field angle of $\SI{73}{\degree}$ with respect to the Si-nanowire axis. Throughout our whole study, the field angle is kept at this value to deliberately set up a nearly symmetric configuration for $g_R^*$ and $g_L^*$, facilitating the interpretation of the results. Small deviations from a fully symmetric situation can be noticed in Fig.~\ref{fig1}~(\textbf{d}) and Fig.~\ref{extracted_gs}.

\subsection{Flopping mode parameters}\label{FM parapmeters}

\begin{figure*}[htbp]
    \includegraphics[width=1\linewidth]{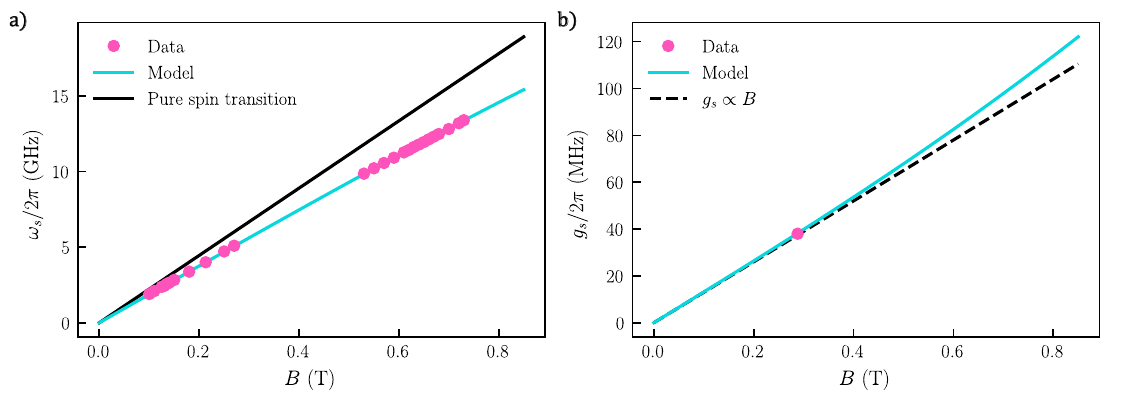}

   \caption{\textbf{Flopping mode model} (\textbf{a}) Measured  magnetic-field dependence of $f_\mathrm{qubit}$ and (\textbf{b}) Spin-photon coupling strength at resonance, fitted together to obtain $\theta$, with $t_c$ and $g_{L,R}^*$ measured beforehand. In this example, the g-factors used in the model ($g_L^*=g_R^*=1.59$) had to be slightly adjusted from the ones measured in single QD regime ($g_L^*=1.62$, $g_R^*=1.67$), which is likely a consequence of a variation of the g-factors with gate-voltage not captured the model. The solid black line in (\textbf{a}) is the expected energy in absence of SOI (Zeeman energy for a spin with g-factor $g=1.59$). The spin-photon coupling is expected to scale linearly with the magnetic field, similar to a spin-like transition in a spin-orbit field, which is a consequence of the relatively weak spin-charge hybridization (e.g $f_\mathrm{qubit} \ll 2t_c$) \cite{Yu}. }
    \label{floppy_model}
\end{figure*}

As explained above, the degree of spin-charge mixing can be determined by fitting the FM qubit energy dependence on magnetic field magnitude. Figure~\ref{floppy_model}(a) shows such a fit with the tunnel coupling and site-specific g-factors constrained by the experimental data of Sect.~\ref{gfactors} and Sect.~\ref{chargeq}. Moreover, with knowledge of the charge-photon coupling (see Sect.~\ref{chargeq}), it is possible to compute the spin-photon coupling for the FM spin qubit at every magnetic field~\cite{Yu} as shown in Figure~\ref{floppy_model}(b). The validity of the modeling is confirmed by the vaccum Rabi splitting shown in the first figure of the main text which reveals a spin-photon coupling $g_s$ in agreement with the model, see Figure~\ref{floppy_model}(b).

Eventually, Table.~\ref{tab_FM} is listing the experimental parameters corresponding to the different figures of the manuscript.

\begin{table}[h]
    \caption{\textbf{Flopping mode parameters:} Measured (black) and estimated (orange) parameters corresponding to the figures indicated in the second column. Estimations are obtained by fitting a dataset with the flopping mode model, as opposed to the measured values obtained with a dedicated experiment. The $1^\mathrm{st}$ line corresponds to a first set of measurements, after which a quick power cycle of the fridge without re-wiring, lead to a change in g-factors and angle $\theta$ (second line) without other noticeable effects. This variation is not mentioned in the main text for simplicity, and these two sets of measurements are mentioned as ``cooldown 1" in the main text. The following longer power-cycle of the fridge, with re-wiring, lead to a more noticeable change which required a full re-measurement of the parameters. It corresponds to the $3^\mathrm{rd}$ line, mentioned as ``cooldown 2" in the main text.}
    \label{tab_FM}
    \centering
        \begin{tabular}{|m{1.5cm}|m{5.6cm}|c|c|c|c|c|c|}
    \hline
    Cooldown & Figures & $\alpha$ & $g_c/2\pi$ (MHz) &
    $t_c/h$ (GHz) & $g_L^*$ & $g_R^*$ & $\theta$ (rad) \\
    \hline

    \multirow{2}{*}{1}
    & Fig.~\ref{fig1}a-c, Fig.~\ref{fig3}a, Fig.~\ref{floppy_model}a-b,
      Fig.~\ref{modes}a, Fig.~\ref{stability},
      Fig.~\ref{LZMS}, Fig.~\ref{chic_meas},
      Fig.~\ref{tuning_spin}
    & $0.47$ & $437$ & $22$
    & \textcolor{orange}{$1.59$} (vs 1.62)
    & \textcolor{orange}{$1.59$} (vs 1.67)
    & \textcolor{orange}{$0.39 \times \pi/2$} \\
    
    \cline{2-8}

    & Fig.~\ref{fig3}c, Fig.~\ref{floppy_model}c-d,
      Fig.~\ref{extracted_gs}a-b
    & $0.47$ & $437$ & $22$
    & $1.64$ & $1.77$
    & \textcolor{orange}{$0.48 \times \pi/2$} \\

    \hline

    2
    & Fig.~\ref{fig1}d, Fig.~\ref{extracted_gs}d,f
    & $0.36$ & $315$ & $20$
    & $1.56$ & $1.52$
    & \textcolor{orange}{$0.30 \times \pi/2$} \\

    \hline
\end{tabular}

\end{table}

\subsection{Methodology and data acquisition for time-domain measurements}\label{sec:dataaqui}
\begin{figure*}[htbp]
    \includegraphics[width=0.8\linewidth]{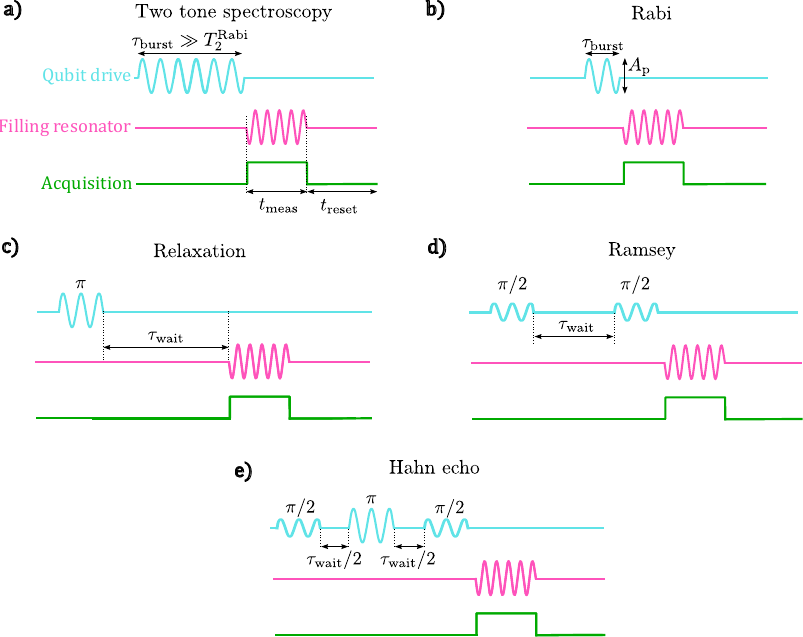}
   \caption{\textbf{Time domain sequences:} Pulse sequences of (\textbf{a}) two-tone spectroscopy showing the sequence to drive the qubit (cyan), fill the resonator (pink), and acquire the output signal (green). (\textbf{b}) Rabi sequence, from which $\pi$ and $\pi/2$ pulses are calibrated. With this pulse calibrations, we measure relaxation (\textbf{c}) and perform Ramsey (\textbf{d}) and Hahn echo (\textbf{e}) experiments by sweeping a waiting time $\tau_\mathrm{wait}$.}
    \label{pulse_sequence}
\end{figure*}

The spin-resonator coupling leads to a dispersive interaction when $|\Delta| \gg g_s$, shifting the resonator frequency by the state dependent ac-Stark shift $\frac{g_s^2}{\Delta} \hat\sigma_z$, allowing to measure the qubit state by measuring the transmission of the resonator \cite{2021_Blais}. All measurements presented in this study are performed in time domain, with the exception of Fig.~\ref{fig1}~(\textbf{c}) and supplementary section \ref{sec:setup}, which are done in continuous wave. Fig.~\ref{pulse_sequence} indicates the pulse sequences used in this study. We typically use $\SI{200}{ns}$ long pulses for the readout at an optimal power of $\sim\SI{-112}{dBm}$ at the chip (assuming a cable loss of $\SI{6}{dB}$), corresponding to around $80$ photons in the resonator. We found that at such a large power, the TWPA provides only a small gain in signal to noise ratio and is therefore left unused. After a reset time $t_\mathrm{reset}$ to let both qubit and resonator relax to their ground states (typically a few microseconds long, largely exceeding $1/\kappa_{r,1}$), this process is repeated around $10^4$ times and averaged to obtain one data point.


We generally performed the following routine to collect the data of Fig.~2, 3 and 4 of the main text:
\begin{enumerate}
    \item Determine $f_\mathrm{qubit}$ in a pulsed two-tone measurement, where the qubit is driven for $\tau_\mathrm{burst}\gg T_2^\mathrm{Rabi}$ to ensure an incoherent mixture, see Fig.~\ref{pulse_sequence}~(\textbf{a}).
    \item Calibrate $\pi$ and $\pi/2$ pulses by applying a resonant drive pulse of $\SI{20}{ns}$ at $f_\mathrm{qubit}$, while varying the pulse amplitude $A_\mathrm{p}$. Fitting the Rabi oscillations as a function of $A_\mathrm{p}$ allows us to extract $A_\pi$, the $\pi$-pulse amplitude, and $A_{\pi/2}\simeq\frac{1}{2}A_{\pi}$, the $\pi/2$-pulses amplitude, see Fig.~\ref{pulse_sequence}~(\textbf{b}).
    \item $T_1$ is extracted by an exponential fit of the cavity signal as a function of $\tau_\mathrm{wait}$ after exciting the qubit by a $\pi$-pulse, see Fig.~\ref{pulse_sequence}~(\textbf{c}).
    \item Dephasing times ($T_\phi^*$ and $T_\phi^\mathrm{e}$) are extracted from the decay envelope of a Ramsey or a Hahn echo sequence, see Fig.~\ref{pulse_sequence}~(\textbf{d}) and (\textbf{e}). We account for the influence of a finite $T_1$, see Sec.~\ref{sec:meas_dephasing},
    
\end{enumerate}

\newpage
\section{Relaxation}
We now turn to the detailed description of the model for relaxation. First we will discuss multimode Purcell  effect followed by a treatment of spin relaxation due to Johnson-Nyquist noise.

\subsection{Multimode Purcell}\label{sec:relax_purcell}
Considering a collection of modes {$m$}, of resonance frequencies {$f_m$} with photon loss rates {$\kappa_m$}, the radiative decay $\Gamma_1$ from a multimode Purcell effect can be written as \cite{Houck_T1}
\begin{equation}
    \Gamma_1 = \sum_m \Gamma_m,
    \label{multimode_Purcell}
\end{equation}
with the Purcell relaxation through mode $m$ given by \cite{2014_Sete}
\begin{equation}
    \Gamma_m = \frac{\kappa_m}{2} \large( 1- \frac{|\Delta_m|}{\sqrt{\Delta_m^2 + 4g_{\mathrm{s}, m}^2}}\large).
    \label{Purcell}
\end{equation}
Here, $g_{\mathrm{s},m}$ is the spin-photon coupling strength of mode $m$ and $\Delta_m = 2\pi (f_\mathrm{qubit}-f_m)$. Equation~\ref{Purcell} assumes that $\kappa_m \ll \sqrt{\Delta_m^2 + 4g_{\mathrm{s},m}}$, which is valid throughout our whole study. At resonance ($\Delta_m=0$), this effect leads to a relaxation peak of $\Gamma_m = \kappa_m/2$ fixed by the losses of the mode, whereas in the dispersive limit ($\Delta_m\gg g_{\mathrm{s},m}$), the relaxation is given by $\Gamma_m \approx \kappa_mg_{\mathrm{s}, m}^2/\Delta_m^2$. The latter can easily be understood by considering that the fraction $g_{\mathrm{s}, m}^2/\Delta_m^2$ of the dressed state is of photonic nature and hence decays with a rate $\kappa_m$. To estimate the overall qubit relaxation through these modes, we therefore need to estimate the modes' frequencies, their photon losses, as well as their coupling to the qubit.

\begin{figure*}[htbp]
    \includegraphics{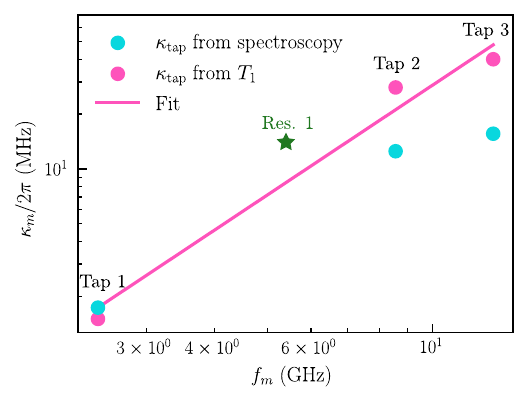}

   \caption{\textbf{Extracting $\kappa_m$}: Photon loss rates of the first three harmonics of the tap extracted by spectroscopy in cyan (see Fig.\ref{modes}),  and by using the lifetime dips in Fig.~3a, governed by $1/T_1 = \kappa_m/2$, as probes of the modes' losses, in pink. The two datasets are in qualitative agreement. Solid line is a fit with eq.~\ref{kappa_scaling} of the data extracted from relaxation. The loss rate of the readout resonator is indicated by a star for comparison.}
    \label{kappa_extraction}
\end{figure*}

The photon-losses $\kappa_m$ can be expressed as the sum of the losses due to the coupling to the feedline $\kappa^c_m$ and the internal losses $\kappa^i_m$, which are for example losses in the dielectric, or losses through other gate lines. From transmission measurements we extracted internal and coupling losses for the fundamental mode of the readout resonator (see Fig.~\ref{modes}~(\textbf{a})). We use the qubit relaxation rates at resonance with the modes of the tap (see Fig.~3a): $1/T_1 \simeq \kappa_m/2$ to directly extract the tap modes photon losses: $\kappa_{\mathrm{tap, 1}}/2\pi = \SI{1.5}{MHz}$, $\kappa_{\mathrm{tap}, 2}/2\pi = \SI{28}{MHz}$ and $\kappa_{\mathrm{tap}, 3}/2\pi = \SI{40}{MHz}$. This increasing dependence with frequency can be captured by the scaling of $\kappa_m$, assuming that  $\kappa^c_m$ scales with $f_m^2$ \cite{2021_Blais} and assuming a frequency independent internal quality factor:

\begin{equation}
    \kappa_m = \kappa_\mathrm{1}^i\frac{f_m}{f_\mathrm{1}} + \kappa^c_{\mathrm{1}} \left( \frac{f_m}{f_\mathrm{1}} \right)^2,
    \label{kappa_scaling}
\end{equation}
where $m=1$ is the fundamental mode. Fitting the data with eq.~\ref{kappa_scaling}, we extract a negligible effective internal loss rate $\kappa_\mathrm{tap,1}^i$ and $\kappa_\mathrm{tap,1}^c/2\pi\sim$\SI{1.7}{MHz} as effective losses to the feedline (see Fig.~\ref{kappa_extraction}). Knowing the internal and coupling losses of the readout resonator (see Fig.~\ref{modes}~(\textbf{a}) ) as well as that of the tap modes (see Fig.~\ref{kappa_extraction}), we use the frequency scaling of $\kappa_m$ (Eq.\ref{kappa_scaling}) to predict the photon losses of higher harmonics that are not measurable in our study (e.g $f_m>\SI{16}{GHz}$).


The third ingredient of Eq.\ref{Purcell}, $g_{\mathrm{s}, m}$ can be expressed according to \cite{Yu}:

\begin{equation}
    g_{s, m} = g_{c,m} |\langle - \uparrow | \tau_z| - \downarrow \rangle|,
    \label{gs}
\end{equation}
where $g_{c,m}$ is the coupling of the charge qubit to mode $m$ at $\varepsilon=0$, which is proportional to the zero-point fluctuation $V_{\mathrm{zpf}, m}$ of mode $m$. The  matrix element $|\langle - \uparrow | \tau_z| - \downarrow \rangle|$ depends on the spin-charge mixing and is the same for all modes. The zero-point fluctuation scales with the frequency according to \cite{2020_Gao}: $V_{\mathrm{zpf}, m} = \sqrt{hf_m/Lc} \propto \sqrt{f_m}$, with $L$ the length of the mode and $c$ the capacitance per unit-length, two quantities which are shared by the resonator and the tap modes. Note that for the fundamental mode of a $\lambda/2$ resonator, $f_1 = (2LZc)^{-1}$ where $Z$ is the impedance of the resonator, leading to the commonly used expression $V_{\mathrm{zpf},1}=f_1\sqrt{2hZ}$ \cite{Yu, 2024_DePalma}.

Thus we express the coupling of any mode to the qubit as:

\begin{equation}
    g_{s, m} = g_s \sqrt{\frac{f_m}{f_\mathrm{r,1}}}.
    \label{gs_modes}
\end{equation}

With the knowledge of $g_s(f_\mathrm{qubit})$, given by the FM model (see Fig.~1 of the main text and section \ref{FM parapmeters}), we can estimate the coupling of the spin qubit to all other harmonic modes generated by the tap and the resonator.

In the following, we discuss which modes we take into account for the multimode Purcell model presented in the main text. By simply considering the resonator and the tap as three equivalent segments of a transmission line, where two ends are open (coupling to feedline and coupling to qubit) and one end is shorted to ground (G2), we would expect a series of $\lambda/4$ modes with frequencies given by $f_{\mathrm{tap}, n} = (2n-1)f_{\mathrm{tap}, 1}$ with $n\in\mathbf{N}$ and a series of $\lambda/2$ modes, where only odd harmonics are present leading to $f_{\mathrm{r}, n} = (2n-1)f_{\mathrm{r}, 1}$ with $n\in\mathbf{N}$ \cite{CollardFiltering}. The absence of the even modes in the $\lambda/2$-spectrum is confirmed by the absence of a relaxation dip at $\sim\SI{10.8}{GHz}$. Summing these harmonics up to infinity would lead to a diverging relaxation rate \cite{Houck_T1}, we therefore only consider the modes below a cut-off frequency of $\SI{200}{GHz}$, which is chosen so that the resulting multimode Purcell relaxation matches the $B^{-2}$ background in Fig.~\ref{fig3}~(\textbf{a}).

For modes with frequencies $f_m \gg f_\mathrm{qubit}$, $\Delta_m$ can be approximated by  $2\pi f_\mathrm{qubit}$ and the Purcell relaxation rate reduces to $\Gamma_m \approx \kappa_m (g_{s, m}/2\pi f_m)^2$. Here, the spin-photon coupling is the only term with a magnetic field dependence, which leads to $\Gamma_m \propto B^{2}$, reproducing the magnetic field dependence in between the resonant dips in Fig.~\ref{fig3}~(\textbf{a}). It is equivalent to relaxing to a photonic bath with a constant density of state, created by modes at much higher frequencies than $f_\mathrm{qubit}$.

The divergence of the multimode Purcell effect is well known in literature \cite{2021_Blais, 2017_Malekakhlagh, 2017_Gely} and arises from a too simplistic treatment of the problem. Finite capacitances of the qubit itself leads to a decrease in $g_{c, m}$ for large frequencies. This can simply be understood by the fact that the presence of the qubit introduces a shunt capacitance of the resonator end to ground. This transforms the open end of the resonator (e.g. voltage anti-node) into a shorted end (e.g. voltage node) for frequencies where the impedance of that capacitance becomes much smaller than the characteristic impedance of the resonator. Therefore, the electic-dipole coupling vanishes. In our case, this shunt capacitance is the capacitance of the resonator to all surrounding electrodes, which are grounded at high frequency through their large on-chip filter capacitance ($\sim\SI{0.134}{pF}$). A cut-off at \SI{200}{GHz} will require a capacitance on the order of \SI{0.3}{fF}, which is in qualitative agreement with \SI{0.7}{fF} extracted from finite element simulations using Sonnet.



\subsection{Johnson-Nyquist}\label{sec:relax_JN}

An impedance $Z$ at temperature $T$ generates voltage fluctuations whose quantum spectral density follows \cite{1997_Devoret}: 

\begin{equation}
    \begin{split}
        S_V(\omega) =  \int dt\langle V(0)V(t)\rangle e^{-i\omega t} \\
        =\hbar \omega\left(\coth\left(\frac{\hbar \omega}{2k_BT}\right)+1\right)\Re\left[Z(\omega)\right]
    \end{split}
\end{equation}
with $V$ the voltage fluctuating across $Z$. Following reference \cite{2005_Ithier}, the Johnson Nyquist (JN) noise leads to a depolarization rate :

\begin{equation}
    \begin{split}
        \frac{1}{T_1}=d^2\frac{S_V(\omega)+S_V(-\omega)}{4} \\
         = \frac{d^2 \hbar \omega}{2} \coth\left(\frac{\hbar \omega}{2k_BT}\right)\Re\left[Z(\omega)\right],
    \end{split}
    \label{JN}
\end{equation}
with $\omega = 2\pi f_\mathrm{qubit}$ and $d$ the electric susceptibility of the spin. $d$ can be estimated from the frequency of Rabi oscillation triggered by a resonant drive of amplitude $V_d$: $2\pi f_\mathrm{Rabi} = d \cdot V_d$. As the zero point voltage fluctuations of the resonator drives the qubit at a Rabi frequency $g_s/\pi$, we estimate $d$ using $d =2g_s/V_\mathrm{zpf}$. The impedance of the resonator is obtained from its design: $Z_r \simeq\SI{2.5}{k\Omega}$ rendering $V_\mathrm{zpf}=f_{\mathrm{r}, 1}\sqrt{2hZ_r} \sim \SI{10}{\micro V}$, and $g_s$ is given by the flopping-mode model for all magnetic fields.

The resulting relaxation time for an impedance $Z=\SI{460}{\Omega}$ at $T=\SI{200}{mK}$ is represented in Fig.~\ref{fig3}~(\textbf{a, c}) (orange curves). The apparent $B^{-2}$ scaling of the JN relaxation at low frequencies comes from the fact the $T_1\propto d^{-2}$ where $d\sim g_\mathrm{s}\sim B$ in the range investigated here (see also Fig.~\ref{fig1}~(\textbf{b}) of the main text for $g_\mathbf{s}(B)$). This regime corresponds to stimulated emission of the spin qubit into a thermally populated electromagnetic environment (equipartition theorem). The crossover towards $B^{-3}$ at $\omega = k_\mathbf{B}T/h$ is due to the linear increase of density of states of electromagnetic modes with increasing energy (here $B$). 

\subsection{Extracting $g_s$ from $T_1$}
\begin{figure*}[htbp]
    \includegraphics[width=1\linewidth]{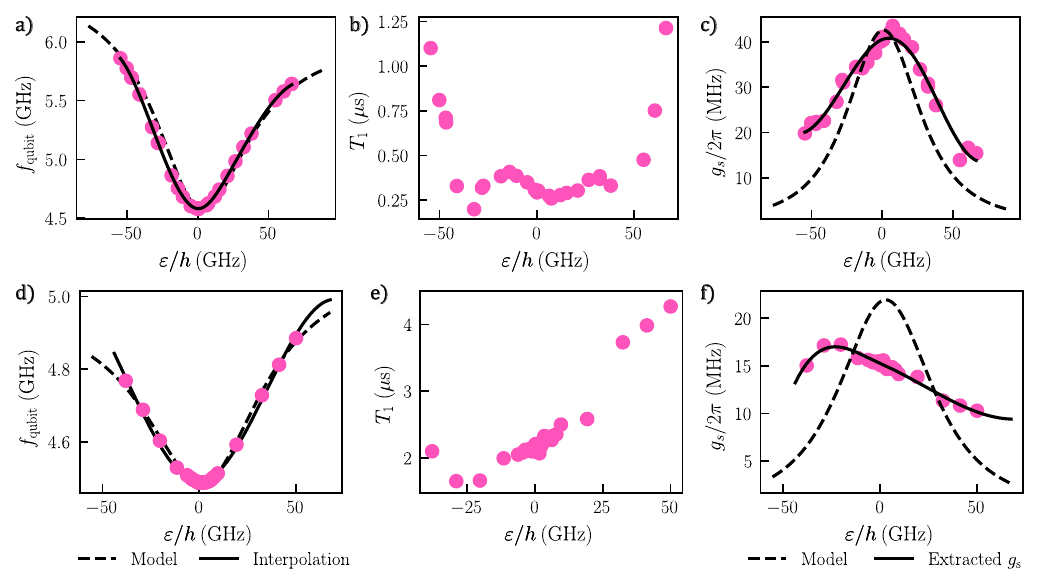}

   \caption{\textbf{$f_{\mathrm{qubit}}$, $T_1$ and extracting $g_s$} (\textbf{a}) $f_\mathrm{qubit}$ extracted from 2-tone measurements and used to drive the qubit in Fig.~4~a)~b). (\textbf{b}) relaxation time, used to obtain the dephasing times in Fig.~4~a)~b) according to Measuring dephasing. The prediction from FM model (dashed line) in a) is in excellent agreement with the data. The solid line is an interpolation of $f_\mathrm{qubit}$ combined with the relaxation times of b) to extract $g_s$'s using Eq. \ref{extracted_gs}. The resulting spin-photon coupling, plotted in \textbf{c)}, is in qualitative agreement with the coupling estimated with the FM model (dashed line). The solid line corresponds to the values of $g_s$ retained to model spin-photon coupling in Photon-Induced dephasing. The same is procedure is reproduced in \textbf{d}), \textbf{e}) and \textbf{f}) with the frequencies and relaxation measured while measuring the dephasing times of Fig.~\ref{fig4}~c)~d).}
    \label{extracted_gs}
\end{figure*}

Assuming that the decay is entirely radiative, the relaxation when the qubit is in the range of \SIrange{4}{6}{GHz} (e.g in vicinity of $f_{r,1}$) can be expressed by:

\begin{equation}
    \frac{1}{T_1}=\Gamma_{\mathrm{r},1} + \left(\frac{g_s}{2\pi}\right)^2 C
\end{equation}
where $\Gamma_{\mathrm{r},1}$ is the Purcell relaxation through the resonator ($f_{\mathrm{r},1}$), and $C$ gathers the effects causing the $B^{-2}$ background of Fig.~1~(\textbf{a}), e.g probably relaxation from other modes and/or Johnson Nyquist. $C$ can be directly estimated from the JN noise needed to reproduce the $B^{-2}$ background, yielding $C=\SI{2.2e-9}{s}$. 

Additionally, assuming a dispersive regime with the resonator leads to $\Gamma_{\mathrm{r},1} = \kappa_{\mathrm{r},1}(g_s/\Delta)^2$, and allows to compute $g_s$ from $T_1$: 
\begin{equation}
    g_s = \frac{1}{\sqrt{T_1\left(\frac{\kappa_{\mathrm{r},1}}{\Delta^2}+\frac{C}{(2\pi)^2}\right)}}
\end{equation}

We apply this in Fig.~\ref{extracted_gs} to extract $g_s$ from $T_1$ measurements, where the dispersive approximation is correct everywhere, with the exception of a few points close to the resonator in Fig.~\ref{extracted_gs}~(\textbf{a}) where $\Delta \sim 5g_s$. Figures~\ref{extracted_gs} (\textbf{a}) and (\textbf{d}) show the measurement of $f_\mathrm{qubit}$ which is well fitted by our FM model. From the measurement of $T_1$ in these configurations, shown in Fig. ~\ref{extracted_gs}~(\textbf{b, e}), we extract the corresponding $g_s$, which are shown in Fig. ~\ref{extracted_gs}~(\textbf{c, f}). While the extracted $g_\mathrm{s}$ and the one predicted from the FM model are pretty close in (\textbf{c}), a larger discrepancy is observed in (\textbf{c}). The FM model generally underestimates $g_\mathrm{s}$ at large $\varepsilon$, as it does not capture the spin-photon coupling in the single quantum dot limit \cite{Yu}. In addition, the FM model used here neglects the variations in $\varepsilon$ of the tunnel-coupling and g-matrices \cite{Yu}. This will lead to variations of the spin-photon coupling terms, which are not necessarily symmetric with respect to $\varepsilon$ and that do not simply evolve with the charge dipole in the DQD. Nevertheless, the spin relaxation time is a viable proxy to infer $g_\mathrm{s}$.

The data presented in Fig.~\ref{extracted_gs} is used to infer the dephasing times presented in Fig.~\ref{fig4} of the main text, where the $T_1$ data is used to extract the dephasing times from coherence measurements (Ramsey and Hahn echo measurements). In addition, we use the extracted $g_\mathrm{s}$ to estimate the dephasing due to photon shot noise (see section~\ref{sec:photon_noise}).

\section{Dephasing}

\subsection{Measuring Dephasing}\label{sec:meas_dephasing}
To measure the Hahn echo and Ramsey dephasing times, we consecutively measure $T_1$ and perform control sequences following Ramsey ($\pi_x/2 - \tau - \pi_\psi/2$) and Hahn echo ($\pi_x/2 - \tau/2 - \pi_y  - \tau/2 - \pi_\psi/2$) sequences, see also Fig.~\ref{pulse_sequence}. 

\begin{figure*}[htbp]
    \includegraphics[width=1\linewidth]{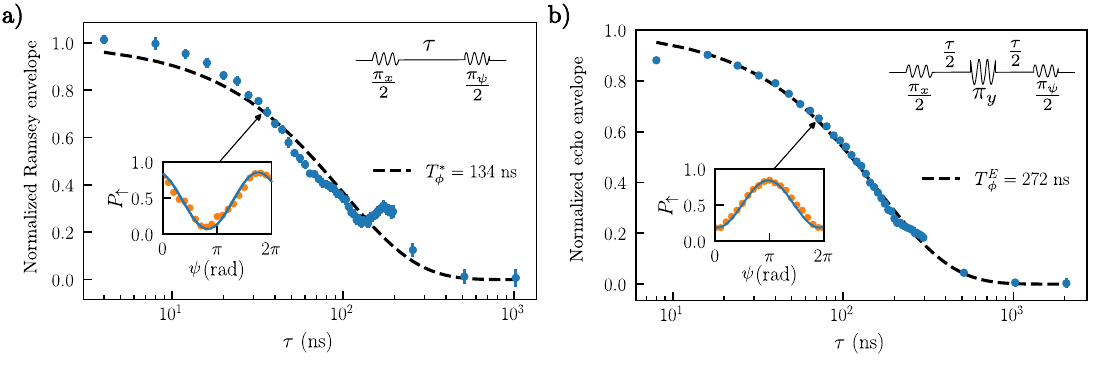}

   \caption{\textbf{Dephasing enveloppes} (\textbf{a}) Ramsey and  (\textbf{b}) Hahn echo envelopes measured at $\varepsilon=0$ and $f_\mathrm{qubit}=\SI{5}{GHz}$. The top right insets depicts the control sequence. The phase $\psi$ of the last pulse is varied over $2\pi$ (see bottom left insets). We extract the amplitude of the oscillation to recover the coherence envelope at each $\tau$. The error bars correspond to the fit uncertainty.}
    \label{dephasing_enveloppes}
\end{figure*}

In order to remove some experimental backgrounds, which can be given by an ill-calibration of the driving frequency for example, we vary the phase $\psi$ of the last $\pi/2$ pulse for both sequences. With perfect pulses, the qubit would do a round-trip around the Bloch-sphere. We recover here a normalized excited-state probability $P_{\uparrow}$, which we fit with a $\sin\left(\psi\right)$ to extract the decaying envelopes (see insets of Fig.\ref{dephasing_enveloppes}). 

Following \cite{2011_Bylander}, the resulting decaying envelopes are fitted with:
\begin{equation}
    P(t) = P_0 \exp\left(-\frac{t}{2T_1}\right) \exp\left(-\left( \frac{t}{T_\phi} \right)^\beta\right)
    \label{Multimode_vs_detuning}
\end{equation}
where imperfections of the pulses are cast into $P_0$. $\beta$ is a decay exponent generally related to the noise color in the case of dominant linear coupling, e.g. first order approximation. In the case of a linear coupling to charge noise (e.g. $1/f$-noise), it is expected that $\beta=2$ \cite{2017_Yoneda, 2005_Ithier} while a coupling to photonic noise (e.g. white noise) would yield $\beta=1$ \cite{2005_Bertet, 2005_Ithier}.

To illustrate our datasets, two decaying enveloppes measured in the first cooldown are displayed on Fig.~\ref{dephasing_enveloppes}. A small beating can be seen in the Ramsey dephasing, which we attribute to a nearby fluctuator seen also in spectroscopy (not provided here), impacting the extraction of $\beta$. As $\beta=1$ works well for both decays, we forced $\beta=1$ to extract the dephasing times in datasets acquired during the first cooldown: Fig.~\ref{fig4}~(\textbf{a-b}) in the main text and Fig.~\ref{photon_shot_noise}). The signatures of this fluctuator disappeared after changing the wiring, so $\beta$ is let free in the fits after (Fig.~\ref{fig2} and Fig.~\ref{fig4}~(\textbf{c-d}) in the main text). 


\subsection{Charge-induced dephasing}\label{sec:charge_noise}
Charge noise is a known limitation for the dephasing time of Si hole spins qubits. Changes in the hole spin wave function due to electrical noise will lead to changes of the g-factor and hence of the Zeeman energy. The qubit response to variations of electric fields can be experimentally probed by measuring the impact of a gate voltage $V_\mathrm{G}$ on the qubit frequency: $\partial f_\mathrm{qubit}/\partial V_\mathrm{G}$ \cite{2022_PiotBrun}. 

\begin{figure*}[httbp]
    \includegraphics[width=1\linewidth]{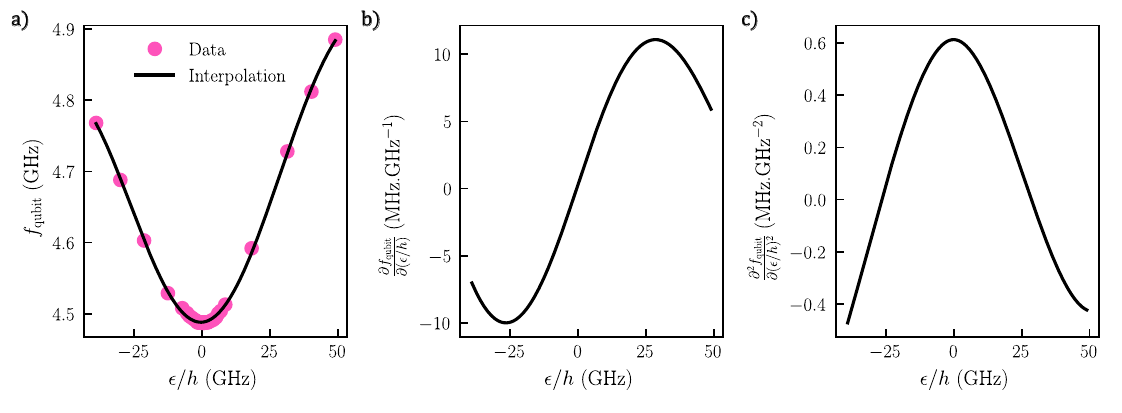}

   \caption{\textbf{Derivatives of $f_\mathrm{qubit}$} (\textbf{a}) Frequencies used to drive the qubit in Fig.~\ref{fig4}~(\textbf{c,~d}). The straight line corresponds to an interpolation, whose first and second order derivatives are plotted on (\textbf{b}) and (\textbf{c}) respectively.}
   
    \label{derivatives}
\end{figure*}

In a DQD, the spin energy strongly depends on the energy detuning $\varepsilon$ between the two dots, as highlighted in Fig.~\ref{derivatives} and Fig.~1~(\textbf{c}) of the main text. The susceptibility to $\varepsilon$-noise, shown in Fig.~\ref{derivatives}~(\textbf{b}), reaches $\partial f_\mathrm{qubit}/\partial \varepsilon = \SI{3}{MHz/\micro eV} $, two orders of magnitude larger than for hole spins in single QDs \cite{2022_PiotBrun}, making $\varepsilon$-noise a relevant source of dephasing.

To study the origin of dephasing, we fit the Ramsey dephasing times of Fig.~\ref{fig4}~(\textbf{a})~and~(\textbf{c}) using the formula for a linear coupling to $\varepsilon$-noise \cite{2005_Ithier}:
\begin{equation}
    \frac{1}{T_{\phi}^*} = 2\pi\left|\frac{\partial f_\mathrm{qubit}}{\partial \varepsilon}\right|\sqrt{A_\varepsilon f_0 \ln \left(\frac{f_\mathrm{uv}}{f_\mathrm{ir}}\right)},
    \label{T2chargeNoiseIthier}
\end{equation}
assuming that the noise follows gaussian statistics with a power spectral density $S_\varepsilon (f)=A_\varepsilon f_0/f$ where $f_0=\SI{1}{Hz}$, and $f_\mathrm{uv}$ (resp. $f_\mathrm{ir}$) the highest (resp. lowest) noise frequency probed during the experiment. The duration of one measurement, typically $\SI{15}{min}$, results in a low frequency cut-off $f_\mathrm{ir}=\SI{1}{mHz}$. For the high-frequency cut-off, we take $f_\mathrm{uv}=\SI{1}{GHz}$, of the order of the qubit frequency. We note that the scaling in $\sqrt{\ln}$ of these frequencies allows us to remain qualitative. The fit of our Ramsey dephasing times yields a noise amplitude of $\sqrt{A_\mathrm{\varepsilon}}$~=~\SI{0.2}{\micro eV /\sqrt{Hz}} for Fig.~\ref{fig4}~(\textbf{a}) and \SI{0.1}{\micro eV /\sqrt{Hz}} for Fig.~\ref{fig4}~(\textbf{c}) in the main text.

With an echo sequence, such gaussian $1/f$ noise is expected to lead to an echo-dephasing time following \cite{2005_Ithier}:

\begin{equation}
    \frac{1}{T_\phi^\mathrm{e}}=2\pi\left|\frac{\partial f_\mathrm{qubit}}{\partial \varepsilon}\right|\sqrt{A_\varepsilon f_0\ln\left(2\right)},
    \label{T2chargeNoiseEcho}
\end{equation}
which we use to fit Fig.~4~(\textbf{b})~and~(\textbf{d}) in the main teyt. The fit yields in both cases$\sqrt{A_\mathrm{\varepsilon}}$~=~\SI{0.1}{\micro eV /\sqrt{Hz}}. We would like to note that the $\varepsilon$-noise amplitudes we extract from the four sub-panels in Fig.~4 of the main text are all in excellent agreement with each other.

At the sweet-spot, where $\partial f_\mathrm{qubit}/\partial \varepsilon=0$, the second-order contribution to dephasing from detuning noise is given by \cite{2005_Ithier}:

\begin{equation}
    \frac{1}{T_\phi^*}=\pi^2\left|\frac{\partial^2f_\mathrm{qubit}}{\partial\varepsilon^2}\right|A_\varepsilon f_0.
    \label{2ndOrderIthier}
\end{equation}

The second derivative of the qubit frequency with respect to $\varepsilon$ is shown in Fig.~\ref{derivatives}~(\textbf{c}). It peaks with $\left|{\partial^2f_\mathrm{qubit}/\partial \varepsilon^2}\right|= \SI{40}{kHz/\micro eV^{2}}$ at the sweet-spot ($\varepsilon=0$). Assuming a coupling to a noise with amplitude $\sqrt{A_\varepsilon}=\SI{0.1}{\micro eV/\sqrt{Hz}}$, Eq.~\ref{2ndOrderIthier} renders $T_\phi^*=$~\SI{300}{\micro s} as expected dephasing time due to the $\varepsilon$ noise at the sweet-spot, far beyond our measured dephasing times.

Such a long charge-limited dephasing time at the sweet-spot is a direct consequence of the low $\varepsilon$-noise, which is at state-of-the-art values for charge noise in spin qubit devices \cite{2024_Elsayed, 2022_PiotBrun, 2025_Hamonic}. While such low levels of charge noise are repeatedly reported for silicon nanowire transistors fabricated by CEA-Leti \cite{2022_PiotBrun, 2025_Hamonic}, substantially higher levels of charge noise ($\sim20$ times higher) are also found \cite{2023_Spence, 2023_Klemt}. More studies are needed to point out the origin of these variations.

\subsection{Photon-induced dephasing}\label{sec:photon_noise}
In the dispersive regime, the photonic population of a cavity $\bar{n}$ displaces the qubit frequency through the ac-Stark shift by $\chi \bar{n}/2\pi$, with $\chi = 2 g_s^2/\Delta$ the dispersive shift per cavity photon. Thermal fluctuations in the photon population of a cavity dispersively coupled to a qubit, therefore, dephase the qubit, following \cite{2007_Clerk}:

\begin{equation}
    \frac{1}{T^{th}_{\phi}}=\frac{\kappa}{2}\Re\left[\sqrt{\left(1+\frac{i\chi}{\kappa}\right)^2 + \frac{4i\chi}{\kappa}\bar{n}}-1\right].
    \label{general_PSN}
\end{equation}

In the strong dispersive regime ($|\chi| \gg \kappa$), the qubit is projected each time a photon passes through the resonator leading to a dephasing maximum $1/T_{\phi}^{th}=\kappa\bar{n}$. In the weak dispersive regime ($|\chi| \ll \kappa$, which corresponds to our measurements), the photon population effectively seen by the qubit is reduced and the dephasing writes as \cite{2012_Rigetti, 2016_Yan}:

\begin{equation}
   \frac{1}{T_{\phi}^{th}} = \chi^2\frac{\bar{n}(\bar{n}+1)}{\kappa}.
  \label{T2_photon_noise}
\end{equation}

All electromagnetic modes coupled to the qubit are, in principle, contributing to this effect. However, due to the increase of $\kappa$ with mode frequency (see Fig.~\ref{kappa_extraction}), and to the exponential decrease of the thermal population with mode frequency, we estimate that only the tap and the resonator fundamental modes contribute significantly to dephasing, which leads to: 

\begin{equation}
    \frac{1}{T^{th}_{\phi, \mathrm{tot}}} = \frac{1}{T_{\phi, \mathrm{r}}^{th}}+\frac{1}{T_{\phi, \mathrm{tap}}^{th}}.
\label{res_tap_photons}
\end{equation}

\begin{figure*}[htbp]
    \includegraphics[]{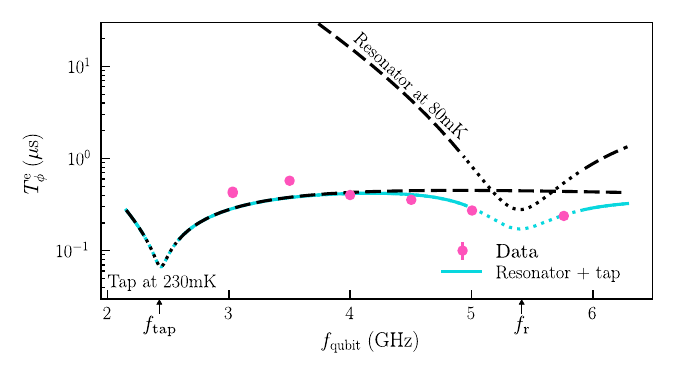}

   \caption{\textbf{Frequency-dependence of $T_\phi^\mathrm{e}$:} $T_\phi^\mathrm{e}$ measured as a function of $f_\mathrm{qubit}$ at $\varepsilon=0$. The data exhibits a decaying trend with frequency which is captured by Eq.\ref{res_tap_photons} assuming a photonic temperature of $\SI{230}{mK}$ for the tap and $\SI{80}{mK}$ for the resonator. Dotted lines corresponds to the regions where the dispersive approximation fails (e.g where $|\Delta_m|<10\cdot g_{s,m}$ for the tap and resonator, respectively). The error bars correspond to the fit uncertainty on $T_\phi^\mathrm{e}$ and are smaller than the data points.}
   
    \label{photon_shot_noise}
\end{figure*}

To extract the photonic temperatures of these two modes, we fit Fig.~\ref{fig3}~(\textbf{b}) of the main text together with a measurement of $T_\phi^\mathrm{e}$ as a function of $f_\mathrm{qubit}$, tuned with $B$, at $\varepsilon=0$, see Fig.~\ref{photon_shot_noise}. The dataset can be captured by Eq.\ref{res_tap_photons}, using the FM model to predict $g_s$ (see Fig.~1~(\textbf{b}) in the main text). We find a tap temperature of $\SI{230}{mK}$ as well as a resonator temperature of $\SI{80}{mK}$ (see cyan line). This measurement also highlights that the resonator has a negligible contribution to dephasing at intermediate frequencies. The fact the dephasing time is asymmetric in energy difference with a given mode comes from the fact that $g_\mathrm{s}$ increases with increasing $f_\mathrm{qubit}$.

To fit the echo-dephasing times versus detuning in Fig.~\ref{fig4}~(\textbf{b}), we use $T_1$ measurements to extract the   $\varepsilon$-dependence of $g_s$ (see Fig.~\ref{extracted_gs}~(\textbf{a-c})), and plot the photon-induced dephasing (PID) predicted from Eq.\ref{res_tap_photons}. The result is captured by an interplay between a PID with photonic temperatures of $\SI{80}{mK}$ for the resonator and $\SI{230}{mK}$ for the tap together with a dephasing caused by charge noise at finite $\varepsilon$. 

To reduce the photonic temperature, we added attenuators on the lines (see Fig.~\ref{device}) and re-measured the dephasing times as a function of detuning (Fig.~\ref{fig4}~(\textbf{d})). The measurements still show signs of being limited by photonic noise at the sweet-spot (see Fig.~\ref{betas}~(\textbf{b})). Taking into account the reduction in $g_s$ (see Fig.\ref{extracted_gs}~(\textbf{d-f})), the observed dephasing is consistent with unchanged photonic temperatures, suggesting that the increased attenuation was ineffective. This indicates that ambient radiation within the fridge may dominate the photonic temperature, rather than radiation from the microwave lines. Similar photonic temperatures due to ambient radiation have previously been reported, emphasizing the need to shield the chip from such radiation \cite{2011_Barends, 2011_Corcoles}.

The temperatures we extract from this fit are only indicative, as the model has many uncertainties, which are: the knowledge of the coupling of the spin to the readout reasonator and to the tap and their variations with $\epsilon$ and $B$, and the value of $\kappa_\mathrm{tap}$. Additionally, Eq. \ref{general_PSN} computes the Ramsey dephasing time while we conveniently extend it to an echo dephasing time. This approximation remains valid as long as $\kappa \gg 1/T_\phi^e$ \cite{2005_Bertet}, and as $\kappa_{\mathrm{tap}, 1}/2\pi = \SI{1.7}{MHz}$, this approximation is valid in Fig.~\ref{fig4}~(\textbf{d}) but not in Fig.~\ref{fig4}~(\textbf{c}).

Nevertheless, there exist explanations for the large difference in the temperatures extracted for the two modes. The temperature of a mode is given by the sum of the temperatures of the bath to which it is coupled to, weighted by the respective coupling rates. Therefore, a mode with a relatively large internal loss rate (e.g. mostly coupled to a cold photonic bath) will have a lower temperature. Given the differences in coupling rates (see supplementary section \ref{sec:relax_purcell}), the resonator mode is expected to be colder than the tap mode.

A second explanation is the likely presence of other spurious modes at lower frequencies in the device environment. This has been observed with a transmon embedded in a box, which was found to couple to the box modes \cite{2021_Huang}. In our case, the puck of the fridge may act as a box with low-frequency fundamental modes. These modes are expected to have weak couplings but large photonic populations. As modes largely detuned in energy from the qubit, they would effectively be captured by the tap mode in our fit, thereby increasing the extracted tap temperature.

\subsection{Detuning dependence of $\beta$}\label{sec:betas}
\begin{figure*}[htbp]
    \includegraphics[width=1\linewidth]{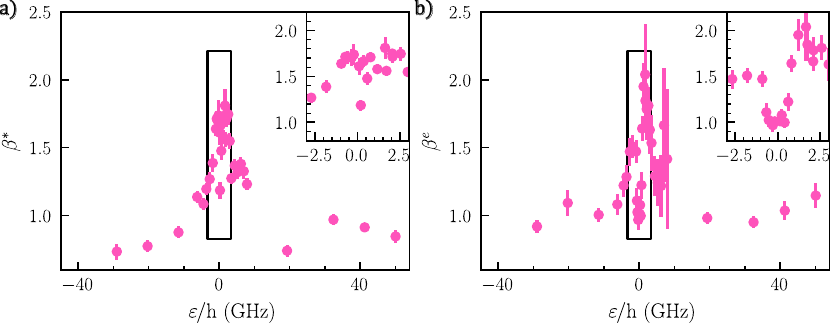}
   \caption{\textbf{Detuning dependence of $\beta$:} decay exponent extracted from the (\textbf{a}) Ramsey  and (\textbf{b}) echo measurements presented in Fig.~\ref{fig4}~(\textbf{c, d}) of the main text. Inset shows a zoom-in of the sweet-spot region highlighted by black rectangles. The error bars correspond to the fit uncertainty.}
    \label{betas}
\end{figure*}

The decay exponents $\beta$ (see eq.~\ref{Multimode_vs_detuning}) of the depahsing times presented in Fig.~\ref{fig4}~(\textbf{c, d}) of the main text are shown in Fig.~\ref{betas}. For the Ramsey experiments, the decay exponent which we refer to as $\beta^*$, reaches $1.7$ at the sweet-spot, close to $2$. For the echo experiments however, the behavior is different at the sweet-spot, where the decay exponent, refereed to as $\beta^e$, sharply drops to $1$ as shown in the inset.

We can link these observations to the noise governing dephasing. Let us introduce the power spectral density on the qubit frequency:

\begin{equation}
        S_{f_\mathrm{qubit}}(\omega) =  \int_{-\infty}^{+\infty} dt \langle f_\mathrm{qubit}(0)f_\mathrm{qubit}(t)\rangle e^{-i\omega t}.
\end{equation}
Assuming a spectra of the form: $S_{f_\mathrm{qubit}}(\omega) = S_0\large(\omega_0/\omega \large)^\alpha$, the decay exponents can be simply expressed as $\beta^* = \beta^e = \alpha+1$, assuming that the qubit frequency has a gaussian noise \cite{2012_Medford, 2022_PiotBrun}. As spin qubit devices are usually subject to gaussian $1/f$ noise ($\alpha=1$), originating from charge or hyperfine fluctuations \cite{BurkardReview}, they typically exhibit gaussian envelopes ($\beta=2$). In Fig.~\ref{betas}~(\textbf{a}), the decay exponent approaching $2$ at the sweet-spot thus indicates a Ramsey dephasing time limited by $1/f$ noise, likely charge or hyperfine noise. This is further supported by observations of slow jumps of the qubit frequency while doing two-tone spectroscopy at the sweet-spot (not provided here), also indicating a slow (e.g $1/f$) noise.

In the case of dephasing induced by the thermal fluctuations of photons in a cavity, the spectra can be expressed in the dispersive regime as \cite{2005_Bertet, 2016_Yan}:

\begin{equation}
        S_{f_\mathrm{qubit}}(\omega) =  \kappa\chi^2 \frac{\bar{n}(\bar{n}+1)}{\omega^2+\kappa^2} \eta,
        \label{photon_spectra}
\end{equation}
with $\eta = \kappa^2/(\kappa^2+\chi^2)\simeq1$ in our case. In consequence, this noise is white ($\alpha=0$) up to $\kappa/2\pi$ (with $\kappa/2\pi \sim \rm{few~MHz}$ for the tap and cavity fundamental modes). 
As an echo sequence probes the noise around $\omega/2\pi\sim 1/2T_\phi^e$, it will have an exponent $\beta^e=1$ if $1/2T_\phi^e$ lies in the white part of the spectrum, e.g if $T_\phi^e>\SI{1}{\micro s}$.

At the sweet-spot of Fig.~\ref{fig4}~(\textbf{d}) we measure $T_\phi^e\sim \SI{5}{\micro s}$, hence the qubit is susceptible to the white part of the photonic noise spectrum. The observation that $\beta^e=1$ at the sweet-spot thus indicates that the dephasing is likely still limited by photons for the echo sequence.

Echo and Ramsey sequences being limited by noises of different nature is a consequence of the fact that they are susceptible to different regions of the noise spectrum (low frequencies for Ramsey and high frequencies for echo~\cite{2011_Bylander}).


\subsection{Rabi oscillation}\label{sec:rabi}
\begin{figure*}[htbp]
    \includegraphics[width=1\linewidth]{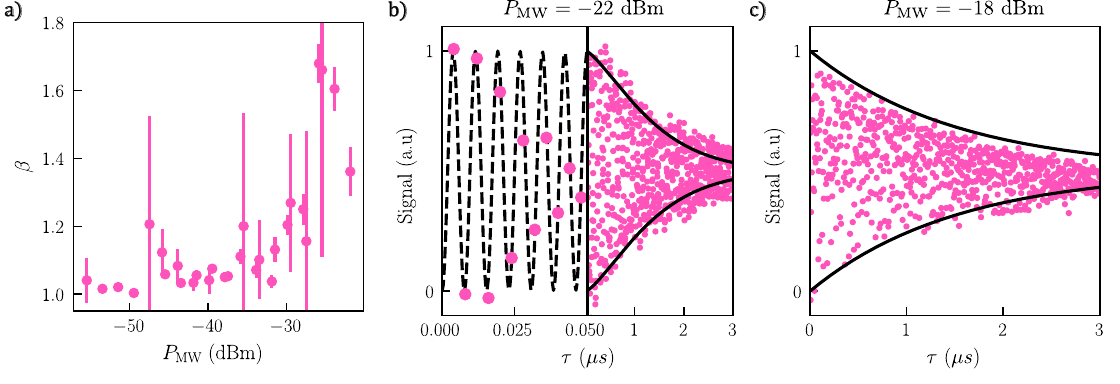}
   \caption{\textbf{Analysis of Rabi oscillations:} (\textbf{a}) decay exponents corresponding to the Rabi oscillations analyzed in Fig~\ref{fig2} of the main text. The error bars represent the uncertainty from the fit of the oscillations. (\textbf{b}) Rabi oscillations at $P_\mathrm{MW}=\SI{-22}{dBm}$. The fitting with Eq.~\ref{rabi_formula} leads to a Rabi frequency of $f_\mathrm{Rabi}=\SI{130}{MHz}$ (dashed line), at the Nyquist limit given by the sampling rate of $1/(\SI{4}{n s})$ (left). Right: envelope of the decay (solid line). (\textbf{c}) Fit of the envelope of a Rabi oscillation at a power of $\SI{-18}{dBm}$  yielding $T_2^\mathrm{Rabi}=\SI{1.5}{\micro s}$.}
    \label{beta_rabi}
\end{figure*}

The analysis of coherent driving reported in Fig.~\ref{fig2} of the main text, is performed by fitting the signal of Rabi oscillations to
\begin{equation}
    P(t) = P_0 \sin(2\pi f_\mathrm{Rabi} t) \exp\left(-\left( \frac{t}{T_2^\mathrm{Rabi}} \right)^\beta\right),
    \label{rabi_formula}
\end{equation}
where the Rabi coherence time $T_2^\mathrm{Rabi}$ captures both dephasing and relaxation effects. $\beta$ allows to reproduce decays of different nature, from pure exponential (caused by photonic noise or relaxation) to gaussian (caused by hyperfine or charge noise). The $\beta$ obtained to produce Fig~\ref{fig2} are shown in Fig.~\ref{beta_rabi}~(\textbf{a}).

The highest gate quality factor reported in Fig~\ref{fig2}~(\textbf{d}) corresponds to a Rabi oscillation at a Rabi frequency reaching the Nyquist limit, see Fig.~\ref{beta_rabi}~(\textbf{b}). In consequence, for higher powers, $f_\mathrm{Rabi}$ cannot be extracted from oscillations. However, the envelopes can still be resolved as shown in Fig.~\ref{beta_rabi}~(\textbf{c}), where extrapolating $f_\mathrm{Rabi}$ from the applied power (solid line in Fig~\ref{fig2}~(\textbf{b})) leads to a quality factor of $Q_\mathrm{gate}=690$. This demonstrates that the gate quality factor reported in the main text is only a lower bound of what can be achieved. At even higher powers, artificial signals in the readout prevent us from fitting the envelopes.

\bibliography{biblio}
\end{document}